\documentclass[manuscript, screen]{acmart}

\AtBeginDocument{%
  }

\setcopyright{acmcopyright}
\copyrightyear{2018}
\acmYear{2018}
\acmDOI{XXXXXXX.XXXXXXX}

\acmPrice{15.00}
\acmISBN{978-1-4503-XXXX-X/18/06}

\usepackage{hyperref}
\usepackage{makecell}
\usepackage{multirow}
\usepackage{booktabs}
\usepackage{subcaption}
\usepackage{graphics}
\usepackage{makecell}
\usepackage{adjustbox}

\usepackage{fontawesome5}
\usepackage{xcolor}

\definecolor{OrangeRed}{RGB}{230, 85, 13}
\definecolor{RoyalBlue}{RGB}{86, 180, 233}
\definecolor{SlateGray}{RGB}{140, 140, 140}
\definecolor{ForestGreen}{RGB}{0, 158, 115}

\definecolor{myred}{RGB}{244, 149, 140}
\definecolor{myblue}{RGB}{112, 164, 255}
\definecolor{mygreen}{RGB}{137, 237, 144}
\definecolor{mygray}{RGB}{216, 216, 216}

\definecolor{darkgray}{RGB}{120, 120, 120}

\newcommand{\A}{{\color{myred} \faAsterisk}}      %
\newcommand{\B}{{\color{myblue} \faCircle[solid]}} %
\newcommand{\C}{{\color{darkgray} \faTimes}}        %
\newcommand{\D}{{\color{black} \faDotCircle[regular]}} %

\newcommand{\GOne}{\textcolor{myred}{\textbf{G1}}}
\newcommand{\GTwo}{\textcolor{myblue}{\textbf{G2}}}
\newcommand{\GThree}{\textcolor{black}{\textbf{G3}}}
\newcommand{\GFour}{\textcolor{darkgray}{\textbf{G4}}}

\usepackage{mwe} %

\newcommand{\ket}[1]{\ensuremath{\left|#1\right\rangle}}
\newcommand{\bra}[1]{\ensuremath{\left\langle#1\right|}}

\newcommand{\norm}[1]{\ensuremath{\left\lVert#1\right\rVert}}
\usepackage{mathtools}

\usepackage{xfp} %
\usepackage{color}
\definecolor{salmon}{rgb}{1.0, 0.55, 0.41}
\definecolor{lime-green}{rgb}{0.2, 0.8, 0.2}
\definecolor{teal-blue}{rgb}{0.21, 0.46, 0.53}

\usepackage[most]{tcolorbox}
\definecolor{qiskitblue}{RGB}{107,159,247}
\definecolor{qiskitpurple}{RGB}{181,135,247}
\definecolor{qiskitgreen}{RGB}{4,180,176}

\newcommand{\code}[1]{\texttt{#1}}

\newtcbox{\badge}[1][red]{
  on line,
  arc=2pt,
  colback=#1!80!black,
  colframe=#1!30!black,
  fontupper=\color{white},
  boxrule=1pt,
  boxsep=0pt,
  left=6pt,
  right=6pt,
  top=2pt,
  bottom=3pt,
}

\usepackage{tikz}
\usetikzlibrary{shapes,arrows}

\usepackage{qcircuit}
\usepackage{subcaption}

\usepackage{listings}

\definecolor{pythonblue}{RGB}{0,0,255}
\definecolor{pythonred}{RGB}{255,0,0}
\definecolor{pythonorange}{RGB}{255,165,0}
\definecolor{pythonpurple}{RGB}{128,0,128}
\definecolor{pythongreen}{RGB}{0,128,0}

\lstdefinelanguage{Python}{
  keywords={False, None, True, and, as, assert, async, await, break, class, continue, def, del, elif, else, except, finally, for, from, global, if, import, in, is, lambda, nonlocal, not, or, pass, raise, return, try, while, with, yield},
  morestring=[b]',
  morestring=[b]",
  comment=[l]{\#},
  commentstyle=\color{pythongreen}\ttfamily,
  stringstyle=\ttfamily
}

\definecolor{celadon}{rgb}{0.67, 0.88, 0.69}
\definecolor{codegreen}{rgb}{0,0.6,0}
\definecolor{codegray}{rgb}{0.5,0.5,0.5}
\definecolor{codepurple}{rgb}{0.58,0,0.82}
\definecolor{backcolour}{rgb}{0.95,0.95,0.92}

\usepackage{xcolor} %

\definecolor{codegreen}{rgb}{0,0.6,0} %
\definecolor{codegray}{rgb}{0.5,0.5,0.5} %
\definecolor{codepurple}{rgb}{0.58,0,0.82} %

\makeatletter
\lst@AddToHook{PreSet}{\normallineskiplimit=0pt}
\makeatother

\lstdefinelanguage{Qasm}{
  basicstyle=\ttfamily,
  keywords={OPENQASM, include, qreg, creg, gate, if, measure, reset, barrier, cx, u3, u2, u1, id, x, y, z, h, s, sdg, t, tdg, rx, ry, rz, cx, cy, cz, swap, ccx, cswap, crx, cry, crz, cu1, cu3, measure, reset, barrier, snapshot, save, load, wait, noise, cnot, cz, swap, iswap, fsim, u, cx, mcx, mcy, mcz, mcu1, mcu3, mcrx, mcry, mcrz, mcswap, mcr, mcs, mct, mcx_gray, mcx_recursive, mcx_vchain},
  keywordstyle=\color{blue}\bfseries,
  ndkeywords={qreg, creg},
  ndkeywordstyle=\color{purple}\bfseries,
  identifierstyle=\color{black},
  sensitive=true,
  comment=[l]{\#},
  commentstyle=\color{green}\ttfamily,
  stringstyle=\color{orange}\ttfamily,
  morestring=[b]',
  morestring=[b]"
}

\usepackage{tcolorbox}
\newcounter{findingCounter}
\begin{document}

\title{A Survey on Testing and Analysis of Quantum Software}

\author{Matteo Paltenghi}
\email{mattepalte@live.it}
\affiliation{%
  \institution{Department of Computer Science, University of Stuttgart}
  \country{Germany}
}

\author{Michael Pradel}
\email{michael@binaervarianz.de}
\affiliation{%
  \institution{Department of Computer Science, University of Stuttgart}
  \country{Germany}}

\renewcommand{\shortauthors}{Paltenghi and Pradel}

\begin{abstract}
  Quantum computing is getting increasing interest from both academia and industry, and the quantum software landscape has been growing rapidly.
  The quantum software stack comprises quantum programs, implementing algorithms, and platforms like IBM Qiskit, Google Cirq, and Microsoft Q\#, enabling their development.
  To ensure the reliability and performance of quantum software, various techniques for testing and analyzing it have been proposed, such as test generation, bug pattern detection, and circuit optimization.
  However, the large amount of work and the fact that work on quantum software is performed by several research communities, make it difficult to get a comprehensive overview of the existing techniques.
  In this work, we provide an extensive survey of the state of the art in testing and analysis of quantum software.
  We discuss literature from several research communities, including quantum computing, software engineering, programming languages, and formal methods.
  Our survey covers a wide range of topics, including expected and unexpected behavior of quantum programs, testing techniques, program analysis approaches,  optimizations, and benchmarks for testing and analyzing quantum software.
  We create novel connections between the discussed topics and present them in an accessible way.
  Finally, we discuss key challenges and open problems to inspire future research.

\end{abstract}



\received{20 February 2007}
\received[revised]{12 March 2009}
\received[accepted]{5 June 2009}

\maketitle

\section{Introduction}

Quantum computing is a new paradigm of computation that promises to find the solution to problems that are intractable for classical computers, or provide a speed up over the classical counterpart, such as factoring large numbers~\cite{shorPolynomialTimeAlgorithmsPrime1999} or searching unsorted databases~\cite{groverFastQuantumMechanical1996}.
The field is attracting increasing attention from researchers, governments, and industry.
While many efforts aim at improving the hardware for quantum computation, there also is a large stream of work on the equally important problem of creating quantum software.
The reliability of this software is crucial, because unreliable or inefficient software can hinder or even nullify progress on the hardware side.

Developing reliable and efficient quantum computing software poses several unique challenges for the developers.
First, unlike classical computers that process bits, which can be either 0 or 1, quantum computers process qubits, which can be in a superposition of 0 and 1.
Besides offering potential to solve new types of problems more efficiently, this leads to a much larger search space when testing possible inputs for quantum programs, stressing the need for efficient testing techniques.
Second, quantum programs use a probabilistic model of computation, where the output of a quantum program is non-deterministic, leading to possibly different outputs for the same input.
This is in contrast to the vast majority of classical programs and makes it harder to check the correctness of a quantum program.
Third, quantum programs exhibit properties typical of quantum mechanics, which are often unintuitive, such as entanglement, superposition, and interference, which can lead to bugs that are hard to find and fix.
Indeed, recent work~\cite{paltenghiBugsQuantumComputing2022,luoComprehensiveStudyBug2022} has documented that the percentage of \emph{quantum-specific bugs} in quantum software constitutes a significant portion ranging from 40\% to 80\% of the total bugs in the software.
Fourth, due to the \emph{no-cloning theorem}~\cite{woottersSingleQuantumCannot1982}, we cannot copy a qubit's state, making it impossible to observe the state without destroying it.
This makes step-by-step debugging of quantum programs extremely challenging, unlike classical programs.

To increase the reliability of the quantum software stack, many testing and analysis techniques have been proposed, including testing approaches to detect bugs in quantum software and program analyses to check its correctness or optimize its performance.
Many optimization techniques lie at the frontier between software and hardware, as they aim to optimize the quantum program for the underlying hardware.
This survey aims to provide a comprehensive overview of the current state of the art in testing and analysis of quantum software, highlighting advancements in the field and open challenges that need to be addressed in the future.

In this survey, we focus on testing and analysis of quantum software, including work that aims at improving either the reliability or the performance of quantum software.
Given the wide range of quantum computing research topics, we focus on software manipulating qubits based on the circuit model, where quantum programs are represented as sequences of quantum gates.
While we acknowledge that simulators can be used for testing quantum programs, we exclude extensive discussion of simulation approaches because their primary goal is to provide an alternative execution environment, not to analyze or test the programs themselves.
In total, our survey covers 102 pieces of work published or publicly shared as pre-print between 2007 and 2023.

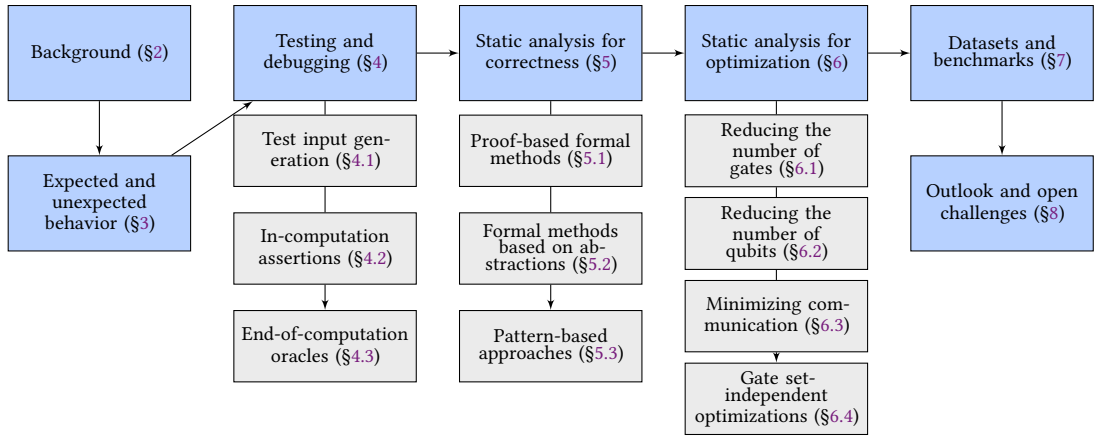
\begin{figure}[t]
  \centering
  \begin{tikzpicture}[node distance=2cm, auto]
    \tikzstyle{block} = [rectangle, draw, fill=lightgray!20,
    text width=7em, text centered, minimum height=4em]

    \tikzstyle{macrochapter} = [rectangle, draw, fill=myblue!50,
    text width=7em, text centered, minimum height=4em, font=\fontsize{7.5}{4.5}\selectfont]
    \tikzstyle{subsection} = [rectangle, draw, fill=mygray!50,
    text width=7em, text centered, minimum height=3em, font=\fontsize{7.5}{4.5}\selectfont]

    \tikzstyle{line} = [draw, -latex']

    \node [macrochapter] (background) {Background (\S\ref{sec:background})};
    \node [macrochapter, below of=background] (behavior) {Expected and unexpected behavior (\S\ref{sec:expected_behavior})};
    \node [macrochapter, right of=background, node distance=3cm] (testing) {Testing and debugging (\S\ref{sec:testing})};
    \node [macrochapter, right of=testing, node distance=3cm] (analysis) {Static analysis for correctness (\S\ref{sec:static_analysis_correct})};
    \node [macrochapter, right of=analysis, node distance=3cm] (optimization) {Static analysis for optimization (\S\ref{sec:static_analysis_optimize})};

    \node [macrochapter, right of=optimization, node distance=3cm] (dataset) {Datasets and benchmarks (\S\ref{sec:dataset})};
    \node [macrochapter, below of=dataset] (implication) {Outlook and open challenges (\S\ref{sec:outlook})};

    \path [line] (background) -- (behavior);
    \path [line] (behavior) -- (testing);
    \path [line] (testing) -- (analysis);
    \path [line] (analysis) -- (optimization);
    \path [line] (optimization) -- (dataset);
    \path [line] (dataset) -- (implication);

    \node [subsection, below of=analysis, node distance=1.3cm] (logic) {Proof-based formal methods (\S\ref{sec:formal_methods_proof_based})};
    \node [subsection, below of=logic, node distance=1.3cm] (simulation) {Formal methods based on abstractions (\S\ref{sec:formal_methods_abstraction})};
    \node [subsection, below of=simulation, node distance=1.3cm] (pattern) {Pattern-based approaches (\S\ref{sec:pattern_based})};

    \path [line] (analysis) -- (logic) -- (simulation) -- (pattern);

    \node [subsection, below of=testing, node distance=1.3cm] (testing_child_1) {Test input generation (\S\ref{sec:testing_input_generation})};
    \node [subsection, below of=testing_child_1, node distance=1.3cm] (testing_child_2) {In-computation assertions (\S\ref{sec:testing_in_computation})};
    \node [subsection, below of=testing_child_2, node distance=1.3cm] (testing_child_3) {End-of-computation oracles (\S\ref{sec:testing_end_of_computation})};

    \path [line] (testing) -- (testing_child_1) -- (testing_child_2) -- (testing_child_3);

    \node [subsection, below of=optimization, node distance=1.3cm] (optimization_child_1) {Reducing the number of gates (\S\ref{sec:g1_reduce_gates})};
    \node [subsection, below of=optimization_child_1, node distance=1.1cm] (optimization_child_2) {Reducing the number of qubits (\S\ref{sec:g2_reduce_qubits})};
    \node [subsection, below of=optimization_child_2, node distance=1.1cm] (optimization_child_3) {Minimizing communication (\S\ref{sec:g3_minimize_communication})};
    \node [subsection, below of=optimization_child_3, node distance=1.1cm] (optimization_child_4) {Gate set-independent optimizations (\S\ref{sec:g4_gate_set_independent_optimizations})};
    \path [line] (optimization) -- (optimization_child_1) -- (optimization_child_2) -- (optimization_child_3) -- (optimization_child_4);

  \end{tikzpicture}
  \caption{Overview of the topics covered in the survey.}
  \label{fig:overview_chapters}
\end{figure}

To contextualize this paper within existing research, we compare it with existing surveys addressing various facets of the field.
Some existing surveys discuss software engineering practices within the quantum computing field~\cite{zhaoQuantumSoftwareEngineering2020}, the various software components within the quantum computing ecosystem~\cite{serranoQuantumSoftwareComponents2022}, and existing quantum programming languages~\cite{heimQuantumProgrammingLanguages2020}.
Other work covers work on benchmarking the hardware components of the quantum computing stack~\cite{reschBenchmarkingQuantumComputers2021}, applications of quantum computing in combinatorial optimization~\cite{gemeinhardtQuantumCombinatorialOptimization2023}, and work on verifying whether a quantum state or operation possesses certain properties~\cite{montanaroSurveyQuantumProperty2016}.
A recent article discusses 16 testing techniques aimed at quantum software~\cite{fortunato2024verification}, i.e., a fraction of those we discuss here.
Overall, despite the large amount of work done on testing and analyzing quantum software, there currently is no comprehensive survey that summarizes this field in a way accessible to interested outsiders.

As shown in Figure~\ref{fig:overview_chapters}, our survey is structured into several sections.
We start by introducing the basic concepts of quantum computing and its software stack (Section~\ref{sec:background}).
Next is a section on expected and unexpected behavior (Section~\ref{sec:expected_behavior}), where we present ways to specify what a quantum program should do.
Following this, Section~\ref{sec:testing} covers different testing techniques and debugging tools.
Subsequently, Section~\ref{sec:static_analysis_correct} and \ref{sec:static_analysis_optimize} discuss work on static analysis.
Specifically, Section~\ref{sec:static_analysis_correct} focused on correctness, i.e., methods that aim to detect bugs in quantum software or prove a program's correctness with respect to some specification.
In contrast, Section~\ref{sec:static_analysis_optimize} focuses on optimization, i.e., methods that aim to optimize quantum software, e.g., by reducing the number of qubits, the number of gates or improving the mapping to the underlying hardware.
We also discuss datasets and benchmarks that are available for evaluating approaches on quantum software testing and analysis (Section~\ref{sec:dataset}).
Finally, we conclude with a discussion of open research challenges and possible directions for future work (Section~\ref{sec:outlook}).

\section{Background}
\label{sec:background}

We introduce essential notation and concepts for this survey.
First, we cover fundamental quantum computing concepts, including the quantum model of computation and state representation.
Next, we discuss the quantum computing software stack, including platform and application code.

\subsection{Quantum Computing Fundamentals}
\begin{figure}[t]
  \centering
  \begin{minipage}[b]{0.18\textwidth}
    \centering
    \subfloat[Quantum circuit. \label{fig:examples_circuit}]{
      \makebox[0.95\textwidth][c]{
        \Qcircuit @C=1em @R=.7em {
          \lstick{q_0}& \gate{H} & \ctrl{1} & \qw \\
          \lstick{q_1}& \qw & \targ & \qw
        }
      }
    }
  \end{minipage}
  \hfill
  \begin{minipage}[b]{0.27\textwidth}
    \centering
    \subfloat[Final state of the circuit in (a). \label{fig:examples_state}]{
      \makebox[0.95\textwidth][c]{
        $\begin{array}{r|c|l}
          $state$ & $prob$ & $bit string$\\
          \begin{pmatrix}
            \frac{1}{\sqrt{2}} \\
            0 \\
            0 \\
            \frac{1}{\sqrt{2}}
          \end{pmatrix} &
          \begin{array}{c}
            50\%\\
            0 \\
            0 \\
            50\%
          \end{array} &
          \begin{array}{l}
            00\\
            01 \\
            10 \\
            11
          \end{array}
        \end{array}$
      }
    }
  \end{minipage}
  \hfill
  \begin{minipage}[b]{0.43\textwidth}
    \centering
    \subfloat[Quantum state as linear combination of basis states. \label{fig:examples_state_linear}]{
      \makebox[0.95\textwidth][c]{
        $\begin{pmatrix}
          \frac{1}{\sqrt{2}} \\
          0 \\
          0 \\
          \frac{1}{\sqrt{2}}
        \end{pmatrix}
        =
        \frac{1}{\sqrt{2}}
        \begin{pmatrix}
          1 \\
          0 \\
          0 \\
          0
        \end{pmatrix}
        + \frac{1}{\sqrt{2}}
        \begin{pmatrix}
          0 \\
          0 \\
          0 \\
          1
        \end{pmatrix}
        =
        \frac{1}{\sqrt{2}} \ket{00} + \frac{1}{\sqrt{2}} \ket{11} $
      }
    }
  \end{minipage}

  \caption{Examples of a quantum circuit (a) that creates an entangled state (b) for which we have 50\% probability to observe the bitstring 00 and 50\% to observe 11. In (c), we show how the output can be represented as a linear combination of the basis states.}
  \label{fig:examples}
\end{figure}
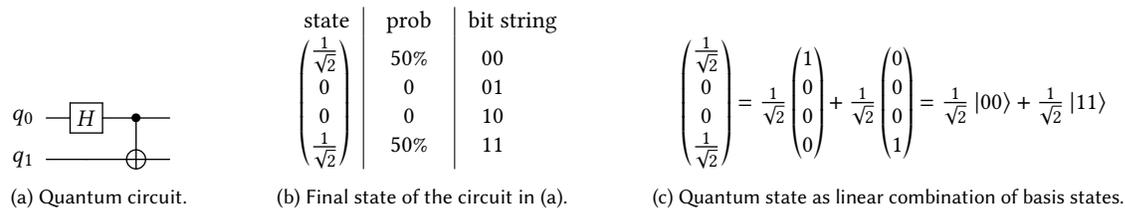

\paragraph{Quantum Model of Computation}
Quantum programs are often represented as circuits, akin to digital logic circuits.
A quantum circuit is a sequence of gates applied to registers containing \emph{qubits}, the fundamental unit of quantum information, similar to classical bits.
Unlike classical bits that are either 0 or 1, qubits can be in both states 0 and 1 simultaneously, known as \emph{superposition}.
Figure~\ref{fig:examples_circuit} shows a quantum circuit, where qubits are lines and gates are boxes.
The computation starts from the left with qubits initialized in a specific state, usually all-zero, and proceeds to the right as qubits are processed by a sequence of gates.

\paragraph{Quantum State}
To allow for superposition, a quantum system's state, comprising one or more qubits, is represented by a column vector of complex numbers.
Each entry, called an \emph{amplitude}, links to the probability of observing a specific classical state upon measurement.
This \emph{state vector}, denoted in Dirac notation $\ket{\psi}$, has length $2^n$, where $n$ is the number of qubits.
The set of all possible states, called \emph{Hilbert space} $\mathcal{H}$, is defined as $\mathcal{H} = \mathbb{C}^{2^n}$, where $\mathbb{C}$ denotes complex numbers.

For a single qubit, its state is a vector of two complex numbers.
In its general form, this state can be written as a superposition of two basis states, $\ket{\psi} = \alpha \ket{0} + \beta \ket{1}$.
Here, $\ket{0}$ and $\ket{1}$ form an orthonormal basis of the Hilbert space~$\mathcal{H}$, known as the \emph{computational basis}.
The coefficients $\alpha$ and $\beta$ are complex numbers whose squared magnitudes give the probability of observing the corresponding basis state, satisfying $\norm{\psi} = 1$.

For multiple qubits, Figure~\ref{fig:examples_state} shows the output state vector of the two-qubit circuit in Figure~\ref{fig:examples_circuit}.
Given the two-qubit circuit, its state vector has length $2^2 = 4$, i.e., $\psi_0 = (\frac{1}{\sqrt{2}}~0~0~\frac{1}{\sqrt{2}})^T$.\footnote{For space reasons, we show the column vectors as transposed row vectors.}
The first qubit $q_0$ is the least significant bit, and the state vector is ordered such that the first complex number corresponds to $\ket{00}$, the second to $\ket{01}$, the third to $\ket{10}$, and the fourth to $\ket{11}$.
Figure~\ref{fig:examples_state_linear} shows the state vector as a linear combination of classical states, where the coefficients are the amplitudes of the bit strings. A classical state is a quantum state where one basis state's amplitude is 1, and the rest are 0.
When two qubits interact, they may become \emph{entangled}, meaning their state cannot be decomposed into individual qubit state vectors.
For instance, in Figure~\ref{fig:examples_circuit}, the final state (Figure~\ref{fig:examples_state}) cannot be decomposed, thus it is entangled.

\begin{figure}[t]
  \centering
  \begin{minipage}[b]{0.30\textwidth}
    \centering
    \subfloat[Application of an H gate to the state $(1~0)^T= \ket{0}$, followed by a measurement. \label{fig:h_gate_measurement}]{
      \makebox[0.95\textwidth][c]{
        \Qcircuit @C=1em @R=.7em {
           & & \lstick{\begin{pmatrix*}[r] 1 \\ 0 \end{pmatrix*}}& \gate{H} & \qw & \ket{\phi} & & \meter \\
        }
      }
    }
  \end{minipage}
  \hfill
  \begin{minipage}[b]{0.26\textwidth}
    \centering
    \subfloat[State vector after applying the H gate (matrix multiplication). \label{fig:h_gate_state}]{
      \makebox[0.95\textwidth][c]{
        \ket{\phi} = $\begin{pmatrix*}[r]
          \frac{1}{\sqrt{2}} & \frac{1}{\sqrt{2}} \\
          \frac{1}{\sqrt{2}} & -\frac{1}{\sqrt{2}} \end{pmatrix*}$
          $\begin{pmatrix*}[r] 1 \\ 0 \end{pmatrix*}$ =
          $\begin{pmatrix*}[r] \frac{1}{\sqrt{2}} \\ \frac{1}{\sqrt{2}} \end{pmatrix*}$
      }
    }
  \end{minipage}
  \hfill
  \begin{minipage}[b]{0.4\textwidth}
    \centering
    \subfloat[Probability of obtaining the state 0 after the measurement. \label{fig:h_gate_measurement_result}]{
      \makebox[0.95\textwidth][c]{
        $p(0)$ = $\begin{pmatrix*}[r] \frac{1}{\sqrt{2}} & \frac{1}{\sqrt{2}} \end{pmatrix*}$
        $\begin{pmatrix*}[r]
          1 & 0 \\ 0 & 0 \end{pmatrix*}$
        $\begin{pmatrix*}[r]
          1 & 0 \\ 0 & 0 \end{pmatrix*}$
        $\begin{pmatrix*}[r] \frac{1}{\sqrt{2}} \\ \frac{1}{\sqrt{2}} \end{pmatrix*} = \frac{1}{2}$
      }
    }
  \end{minipage}

  \caption{Circuit with a Hadamard gate and measurement (a), the corresponding output state after the application of the gate operation (b), and the result of the final measurement (c).}
  \label{fig:gate_and_measurement}
\end{figure}
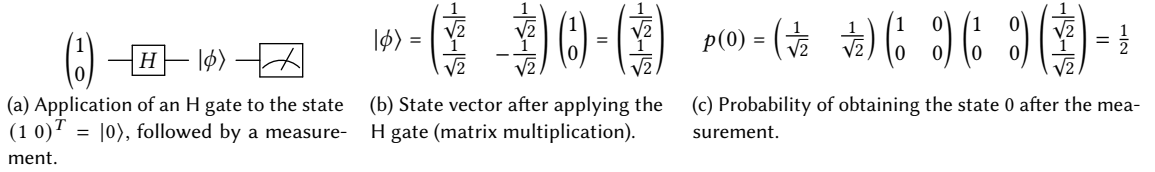

\paragraph{Gates}
Operations applied to qubits are \emph{quantum gates} or \emph{unitary operators}.
A gate transforms the qubits' state vector $G: \mathcal{H} \rightarrow \mathcal{H}$ and is represented by a \emph{unitary} matrix.
A unitary matrix $U$ satisfies $U^\dagger U = I$, where $U^\dagger$ is the conjugate transpose of $U$ and $I$ is the identity matrix.
This ensures that any gate can be inverted by applying its conjugate transpose, making quantum computation reversible.
Quantum computation is a sequence of gates applied to the state vector via matrix multiplication.
For instance, applying a unitary operator $U$ to an initial state $\ket{\psi}_0$ yields $\ket{\psi}_1 = U \ket{\psi}_0$ via matrix-vector multiplication.
The Hadamard gate $H$ is fundamental for creating superposition (Figure~\ref{fig:h_gate_measurement}), with its matrix shown in Figure~\ref{fig:h_gate_state}.
Other notable gates are the single-qubit Pauli gates and the two-qubit CNOT gate.
The Pauli gates ($X$, $Y$, $Z$, $I$) can represent any single-qubit operation when applied sequentially with the right coefficients.
The CNOT gate flips the target qubit if the control qubit is in state $\ket{1}$, and it often entangles qubits, as in Figure~\ref{fig:examples_state}.

\paragraph{Measurement}
Unlike gates, measurement operations are not reversible. A \emph{measurement} is described by $m$ square matrices $\{M_0, ..., M_i, .., M_m\}$ satisfying $\sum_{i=0}^{m} M_i^\dagger M_i = I$, where $i$ indexes possible measurement outcomes.
Applying the operator $M_i$ to the state $\ket{\psi}$ yields outcome $i$ with probability $p(i) = \bra{\psi} M_i^\dagger M_i \ket{\psi}$, and the post-measurement state is $\frac{M_i \ket{\psi}}{\sqrt{p(i)}}$.
For example, a single-qubit measurement in the computational basis, i.e., $\{\ket{0}, \ket{1}\}$, is described by the operators $M_0 = \begin{pmatrix*}[r] 1 & 0 \\ 0 & 0 \end{pmatrix*}$ and $M_1 = \begin{pmatrix*}[r] 0 & 0 \\ 0 & 1 \end{pmatrix*}$.
Figure~\ref{fig:h_gate_measurement_result} shows how to compute the probability of obtaining state $\ket{0}$ after applying the Hadamard gate and measuring the qubit.
Measurement is a probabilistic operation that collapses the quantum state into a classical state.
Once measured, the original quantum state cannot be recovered, but repeating the same measurement on the collapsed state will yield the same result.
The most common measurement uses the Z-basis, related to the Pauli-Z gate, returning either \ket{0} or \ket{1}.
Other measurements, like the X-basis, return \ket{+} or \ket{-}, where \ket{+} = $\frac{1}{\sqrt{2}} (\ket{0} + \ket{1})$ and \ket{-} = $\frac{1}{\sqrt{2}} (\ket{0} - \ket{1})$.

\subsection{Quantum Computing Software Stack}

Quantum computing software is divided into platform code and program code. Platform code provides a mean to define, compile, and execute program code, which in turns solves specific algorithmic problems.

\paragraph{Platform Code}
Referring to \citet{paltenghiBugsQuantumComputing2022}, quantum computing platforms usually provide three main components: (a) a quantum programming language to express programs, (b) a compiler to transform programs before execution, and (c) an execution environment to run programs on simulators or quantum hardware.
Examples include IBM's Qiskit, Google's Cirq, Xanadu's PennyLane, Quantinuum's TKET, and Microsoft's Q\#.
Platforms often integrate analysis and optimization techniques to ensure efficient and correct execution on hardware. Simulators, which precisely track quantum program states, are also useful for debugging aspects unobservable on real hardware.

\begin{figure}[t]
  \centering
  \begin{minipage}{0.65\textwidth}
\centering
\begin{lstlisting}[language=Python, escapechar=|]
from qiskit import QuantumCircuit, Aer, execute
qc = QuantumCircuit(2, 2) |\label{line:qc_creation}|
qc.h(0) |\label{line:qc_h}|
qc.cx(0, 1) |\label{line:qc_cx}|
qc.measure([0, 1], [0, 1]) |\label{line:qc_measure}|
simulator = Aer.get_backend('qasm_simulator')
result = execute(qc, simulator, shots=1000).result() |\label{line:qc_execute}|
print(result.get_counts(qc))
# output: {'00': 502, '11': 498} |\label{line:qc_output}|\end{lstlisting}
  \caption{Qiskit program generating the circuit in Figure~\ref{fig:examples} and performing measurements.}
    \label{fig:example_snippet}
    \end{minipage}
\hfill
  \begin{minipage}{0.25\textwidth}
\centering
\begin{lstlisting}[language=Python, escapechar=|]
q_after:=H(q_before);
\end{lstlisting}
      \vfill
      \vspace*{2em}
        \makebox[0.95\textwidth][c]{
          \Qcircuit @C=1em @R=.7em {
           & & \lstick{\ket{q_{before}}} & \gate{H} & \qw &  & \ket{q_{after}} &  \\
          }
        }
  \caption{Application of Hadamard gate in Silq (top) and its circuit representation (bottom)}
  \label{fig:example_snippet_silq}
  \end{minipage}
\end{figure}

\paragraph{Program Code}
Program code is written by users to solve specific problems using the platform's infrastructure.
Figure~\ref{fig:example_snippet} shows a Qiskit program with a quantum register of two qubits and a classical register of two bits.
Quantum circuits in Qiskit apply a sequence of gates (lines~\ref{line:qc_h}-\ref{line:qc_cx}) to an initial all-zero state, resulting in a final state that is measured (line~\ref{line:qc_measure}).
Since measurement is probabilistic and destructive, the program runs multiple times (line~\ref{line:qc_execute}) to derive a probability distribution reflecting the computation's outcome.
The comment at line~\ref{line:qc_output} underscores the probabilistic nature of quantum computing, showing roughly equal likelihood for bit strings \code{00} and \code{11}.

The representation of quantum gate computation in a program is crucial for analysis.
There are two approaches: \emph{memory-based} and \emph{value-based} semantics.
Memory-based semantics apply qubit operations by referencing the quantum register index, e.g., \code{qc.h(0)} (line~\ref{line:qc_h}), which applies a Hadamard gate to the first qubit.
Developers must track qubit memory locations, similar to pointers in classical programming.
This representation is used by frameworks like Qiskit, Cirq, and the OpenQASM language~\cite{crossOpenQuantumAssembly2017,crossOpenQASMBroaderDeeper2022}.
Instead, value-based semantics represents operations as function calls that take an input quantum state variable and return an output state.
For example, Figure~\ref{fig:example_snippet_silq} shows a Hadamard gate in the Silq language~\cite{bichselSilqHighlevelQuantum2020}.
Besides Silq, this representation is also used by the intermediate representations QSSA~\cite{peduriQSSASSAbasedIR2022} and QIRO~\cite{ittahQIROStaticSingle2022}, which are dialects of the Multi-Level Intermediate Representation (MLIR)~\cite{lattnerMLIRCompilerInfrastructure2020}.

\section{Expected and Unexpected Behavior in Quantum Software}
\label{sec:expected_behavior}

The goal of testing and many analyses is to check that the program behaves as expected.
The following discusses how to specify what is the expected behavior of a quantum program.

\subsection{Specifying Expected Behavior of Quantum Programs}
Specifying the expected behavior of a quantum program is challenging due to its probabilistic nature.
We present three methods: distribution-based, quantum state-based, and unitary-based specifications.

\subsubsection{Distribution-Based Program Specification}
Since the output of a quantum program is probabilistic, the expected behavior of a quantum program is often represented by a probability distribution~\cite{aliAssessingEffectivenessInput2021,wangGeneratingFailingTest2021, wangApplicationCombinatorialTesting2021}.

\begin{definition}[Distribution-based program specification]
  Given a quantum program $P$ with $n$ qubits, a distribution-based program specification $S_\mathit{distr}$ states that, for a valid input $x$, the output follows a probability distribution over the possible outputs of $P$:
    $
        S_\mathit{distr}(P, x) = \{ (y, p) \mid y \in \{0, 1\}^n, p \in [0, 1] \}
    $
    where $y$ is a possible output of $P$ and $p$ is the probability of $y$ being the output of $P$.
    Note that $\sum_{(y, p) \in S_\mathit{distr}(P, x)} p = 1$, and $S_\mathit{distr}(P, x)$ is a set of size $2^n$ because there are $2^n$ possible classical outputs that can be observed with a measurement.
\end{definition}

\subsubsection{Quantum State-Based Program Specification}

Another way to describe the behavior of a quantum program is via the Dirac notation of quantum mechanics~\cite{zhouCoqQFoundationalVerification2023}.
The output quantum state can be described as a superposition of the basis states of the qubits.

\begin{definition}[Quantum state-based program specification]
  Given a quantum program $P$ with $n$ qubits, a quantum state-based program specification $S_\mathit{state}$ states that, for a valid input $x$, the output of $P$ can be represented using either Dirac notation or a vector in Hilbert space:
    $
      S_\mathit{state}(P, x) = \alpha_0 \ket{0} + \alpha_1 \ket{1} + \cdots + \alpha_{2^n - 1} \ket{2^n - 1} = \begin{pmatrix*}[c] \alpha_0 & \alpha_1 & \cdots & \alpha_{2^n - 1} \end{pmatrix*}^T
        $
        where $\alpha_k \in \mathbb{C}$ are amplitudes and $\ket{k}$ are the basis states of the $n$ qubits.
        Note that $\sum_{k = 0}^{2^n - 1} |\alpha_k|^2 = 1$ because the quantum state is normalized.
        As is common practice in quantum mechanics, we represent a possible output in decimal instead of binary, e.g., $\ket{0} = \ket{00\cdots0}$, $\ket{1} = \ket{00\cdots1}$, and $\ket{2} = \ket{00\cdots10}$.
\end{definition}

Using Dirac notation allows applying linear algebra rules to simplify specifications.
For example, the quantum state of three qubits $q_1, q_2, q_3$, represented as $\ket{\psi} = \frac{1}{2} \ket{000} + \frac{1}{2} \ket{010} + \frac{1}{2} \ket{100} + \frac{1}{2} \ket{110}$, can be rewritten as a tensor product of simpler states: $\ket{\psi} = \frac{1}{2} (\ket{00} + \ket{01} + \ket{10} + \ket{11}) \otimes \ket{0}$.
Since the last qubit is always $\ket{0}$, it can be factored out of the tensor product.
As a special case of a quantum state-based program specification, the coefficients can be functions of the input value $x$. For example, the expected output state for the Quantum Fourier Transform (QFT), analogous to the classical Fourier Transform, is:
$S_\mathit{state}(QFT, x) = \frac{1}{\sqrt{2^n}} \sum_{k = 0}^{2^n - 1} e^{2 \pi i k x / 2^n} \ket{k}$
where $x$ is the input, $i$ is the imaginary unit, and $k$ is the basis state index $\ket{k}$.
Distribution-based and quantum state-based specifications are related since the former derives from the latter by squaring the amplitudes' absolute values. Generally, $S_\mathit{state}$ holds more information than $S_\mathit{distr}$, as it can describe negative amplitudes.
For instance, the superposition states $\ket{+} = \frac{1}{\sqrt{2}} (\ket{0} + \ket{1})$ and $\ket{-} = \frac{1}{\sqrt{2}} (\ket{0} - \ket{1})$ are equivalent in a distribution-based specification, both yielding $\ket{0}$ and $\ket{1}$ with probability $\frac{1}{2}$. However, they differ in a quantum state-based specification due to their distinct amplitudes.

\subsubsection{Unitary-Based Program Specification}
A third way to describe the expected output state of a program leverages the representation of quantum operations as matrices~\cite{xuQuartzSuperoptimizationQuantum2022}.
The idea is to specify the expected quantum state as a unitary matrix $U_P$ that transforms an initial all-zero state into the expected output state.
Typically, this matrix is derived by multiplying unitary matrices of simpler operations. This coincides with having a correct reference program $Q$, composed of $n$ known elementary operations, used to compute the expected matrix $U_P$.

\begin{definition}[Unitary-based program specification]
  Given a quantum program $P$ with $n$ gates, its unitary-based specification $S_\mathit{unit}$ is the unitary matrix representing the program:
    $
      S_\mathit{unit}(P) = U_P = U_n \cdot U_{n - 1} \cdot ... \cdot U_1
        $
        where $U_i$ is the unitary matrix of the $i$-th gate in the known program $Q$.
\end{definition}
While each program corresponds to a unique unitary matrix, the same matrix, and thus behavior, can result from different sequences of operations.

\subsection{Specifying Expected Behavior of Quantum Platforms}

Unlike quantum programs that operate on quantum states, quantum computing platforms process classical data, such as source code and intermediate representations.
The expected behavior of a platform is to transform a program while preserving its semantics.
Such transformations convert the program $P_1$, a sequence of $n$ quantum operations, into an equivalent program $P_2$ that is more efficient or hardware-compatible, possibly with a different number of operations.
A common formalism to assess transformation correctness is the \emph{equivalence relation} between two quantum programs.

\begin{definition}[Program equivalence]
    Two quantum programs $P_1$ and $P_2$, represented by two unitary matrices $U_1$ and $U_2$, are equivalent, denoted $U_1 \equiv U_2$, if there exists a real number $\phi \in \mathbb{R}$ such that $U_1 = e^{i \phi} U_2$.
\end{definition}

Note that a special case of program equivalence is \emph{program equality}, denoted $U_1 = U_2$, which is equivalent to $U_1 \equiv U_2$ with $\phi = 0$.
The value $\phi$ is called the \emph{global phase}, a common concept in quantum mechanics where the global phase of a quantum state is not observable. Thus, when measuring the state $\ket{\psi'} = e^{i \phi} \ket{\psi}$, the probability of observing a given outcome is the same as for the state $\ket{\psi}$.
Proving two programs are equivalent means showing they have the same unitary matrix, up to a global phase.

\section{Testing and Debugging}
\label{sec:testing}

\newcommand{\statAssertion}{Statistical}
\newcommand{\ancillaAssertion}{Ancilla-based}
\newcommand{\projectiveAssertion}{Projective}

\begin{figure}[t]
  \centering
  \includegraphics[width=.85\linewidth]{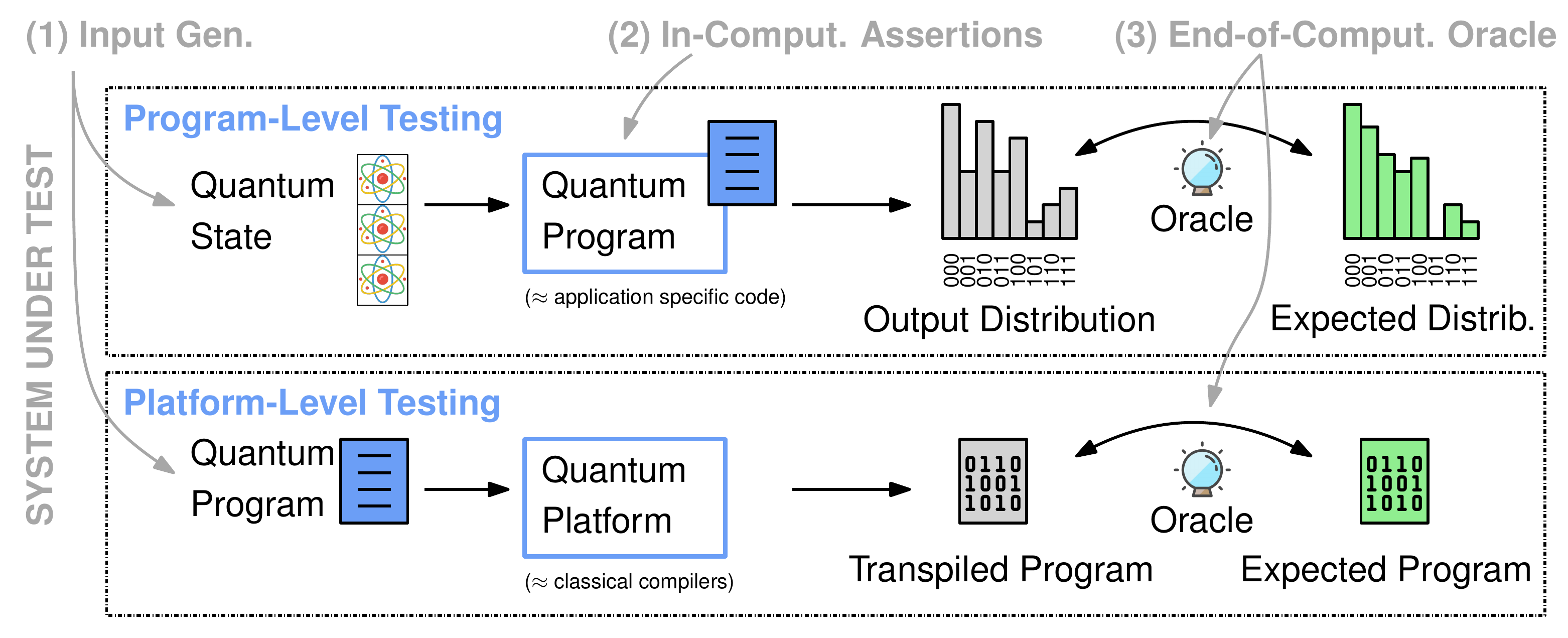}
  \caption{Overview of testing approaches for quantum software.}
  \label{fig:overview_testing}
\end{figure}

Following the literature on software testing, we call the software that gets tested the \emph{system under test} (SUT).
An input given to the SUT, together with a check of the expected behavior of the SUT, form a \emph{test case}.
The input is called the \emph{test input} and the routine that checks if the SUT behaves as expected is called the \emph{oracle}~\cite{barrOracleProblemSoftware2015}.

We group approaches for testing quantum computing software along two axes.
The first axis, shown vertically in Figure~\ref{fig:overview_testing}, concerns the SUT targeted by an approach.
We distinguish between testing quantum programs, akin to traditional application testing, and testing quantum computing platforms, akin to compiler testing~\cite{chenSurveyCompilerTesting2020}.
The second axis, shown horizontally in Figure~\ref{fig:overview_testing}, addresses the testing subproblem tackled by an approach. We consider three subproblems: (1) test input generation, i.e., generating inputs for the SUT; (2) in-computation assertions, i.e., checking program correctness at runtime by asserting properties on a program state, such as the value in a qubit or register; (3) end-of-computation oracles, i.e., verifying the program's output against an expected reference.

Testing quantum programs is uniquely challenging because observing a state destroys it, and copying is impossible due to the no-cloning theorem~\cite{woottersSingleQuantumCannot1982}.
These limitations make debugging harder than for classical programs.
This issue does not affect quantum computing platforms, which adapt classical compiler techniques to quantum theory without performing any quantum execution. Thus, our survey discusses in-computation assertions only for quantum programs.

Sections~\ref{sec:testing_input_generation}--\ref{sec:testing_end_of_computation} detail subproblems in testing quantum software and approaches to address them. Table~\ref{tab:testin_mapping} summarizes the surveyed work. Finally, Section~\ref{sec:debugging} briefly discusses debugging quantum software.

\begin{table}[t]
  \caption{Overview of testing approaches.}
  \centering
  \begin{tabular}{llccc|cc}
  \toprule
  \textbf{Paper} & \textbf{Approach} & \multicolumn{3}{c}{\textbf{Testing subproblem}} & \multicolumn{2}{c}{\textbf{Automation}} \\
  \cmidrule{3-5}
  \cmidrule{6-7}
  & & \textbf{Input gen.} & \multicolumn{2}{c}{\textbf{Oracle}} & \textbf{Manual} & \textbf{Auto}\\
  \cmidrule{4-5}
  &&& \textbf{In-comp.} & \textbf{End-of-comp.} \\
  \midrule
  \multicolumn{3}{l}{\textit{Testing of quantum programs:}} \\
  \midrule
  \citet{wangApplicationCombinatorialTesting2021} &  QuCat & \checkmark & & \checkmark & & \checkmark \\
  \citet{wangGeneratingFailingTest2021} & QuSBT & \checkmark & & \checkmark & & \checkmark \\
  \citet{longTestingQuantumPrograms2023} & QSharpTester & \checkmark & & \checkmark & & \checkmark \\
  \citet{huangStatisticalAssertionsValidating2019} & Stat. Assert. & & \checkmark & \checkmark & \checkmark & \\
  \citet{liuQuantumCircuitsDynamic2020} & Runtime Assert. & & \checkmark & \checkmark & \checkmark & \\
  \citet{liProjectionbasedRuntimeAssertions2020} & Proq & & \checkmark & \checkmark & \checkmark & \\
  \citet{liuSystematicApproachesPrecise2021} & NDD/Swap-based & & \checkmark & \checkmark & \checkmark & \\
  \midrule
  \multicolumn{3}{l}{\textit{Testing of quantum commputing platforms:}} \\
  \midrule
  \citet{paltenghiMorphQMetamorphicTesting2023} & MorphQ & \checkmark & & \checkmark & & \checkmark \\
  \citet{wangQDiffDifferentialTesting2021} & QDiff & \checkmark & & \checkmark & & \checkmark \\
  \citet{xiaFuzz4AllUniversalFuzzing2024} & Fuzz4All & \checkmark & & \checkmark & & \checkmark \\
  \bottomrule
  \end{tabular}
  \label{tab:testin_mapping}
\end{table}

\subsection{Test Input Generation}
\label{sec:testing_input_generation}

Knowing which input to feed to the software under test is crucial for finding bugs.
The problem of generating meaningful test inputs is referred to as test input generation or fuzzing~\cite{zhuFuzzingSurveyRoadmap2022}.

A quantum program with $n$ qubits has $2^n$ possible classical inputs. Considering quantum states as inputs, the number of possibilities becomes virtually infinite due to arbitrary superpositions of those $2^n$ classical inputs. For example, a quantum circuit with $n=2$ qubits has $2^2 = 4$ classical inputs (00, 01, 10, 11), but an infinite number of quantum inputs, such as the superposition state $\ket{\psi} = \frac{1}{\sqrt{2}} (\ket{00} + \ket{11})$.
This vast input space makes exhaustive testing impractical, so techniques to find likely bug-inducing inputs have been proposed. For instance, QuCat~\cite{wangApplicationCombinatorialTesting2021} uses combinatorial testing~\cite{nieSurveyCombinatorialTesting2011} to generate test inputs for quantum programs, similar to classical software testing.
Similarly, QuSBT~\cite{wangGeneratingFailingTest2021} employs genetic algorithms for search-based software testing~\cite{harmanSearchbasedSoftwareEngineering2012} to generate test inputs that maximize failing test cases. While effective on 30 programs with injected bugs, QuSBT has yet to be applied to real-world programs.
To reduce the burden of the large input space, \citet{longTestingQuantumPrograms2023} propose QSharpTester, generating test inputs based on equivalence classes, including classical and superposition states.

In testing quantum platforms, various methods generate quantum programs as test inputs to trigger crashes or unexpected output distributions.
Using differential testing~\cite{chenSurveyCompilerTesting2020}, \citet{wangQDiffDifferentialTesting2021} introduce QDiff, which mutates programs by adding, removing, and replacing operations, or by executing the same operation on different qubits.
Using metamorphic testing~\cite{chenMetamorphicTestingNew2020}, \citet{paltenghiMorphQMetamorphicTesting2023} propose MorphQ, which generates valid quantum programs via templates and a grammar-based generator, then applies quantum-specific metamorphic transformations.
A metamorphic transformation consists of a program transformation and an expected effect on the program's semantics, e.g., applying a sub-circuit and its inverse, which is expected to preserve the original program semantics.
These transformations enable MorphQ to explore complex platform features, e.g., advanced optimizations and OpenQASM machine code conversion.
More recently, inspired by the success of large language models, \citet{xiaFuzz4AllUniversalFuzzing2024} introduce Fuzz4All, a versatile fuzzer that generates realistic-looking quantum programs using large language models, effectively uncovering corner cases in the targeted quantum computing platform.
Additionally, Fuzz4All can target specific platform features derived from the platform documentation in natural language, such as the most recent changelog.

\subsection{In-Computation Assertions}
\label{sec:testing_in_computation}

\begin{figure}[t]
  \centering
  \includegraphics[width=\textwidth]{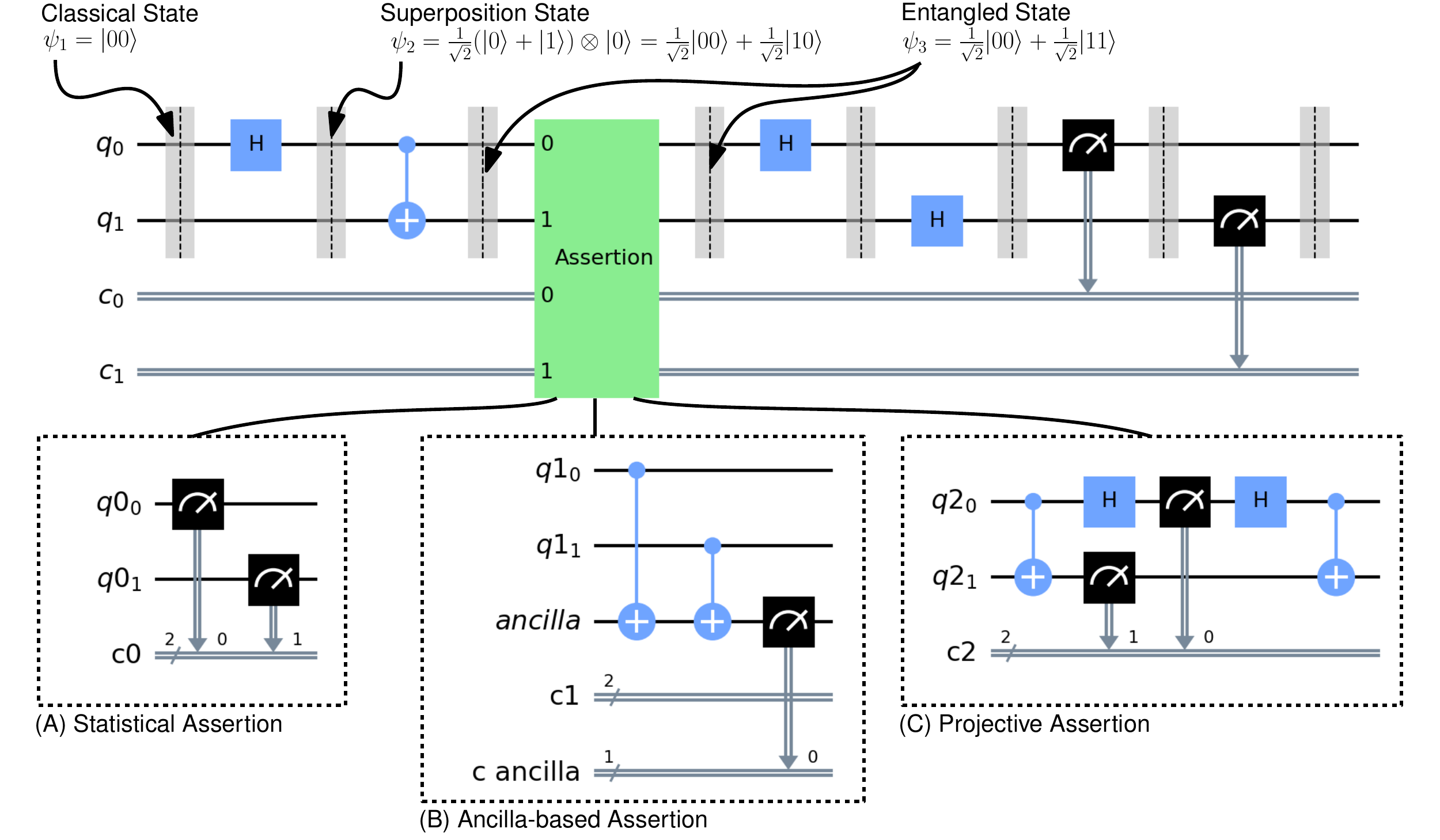}
  \caption{Example circuit under test (top) with three kinds of approaches to check assertions during the computation (bottom).}
  \label{fig:assertion_circuits_for_entangled_states}
\end{figure}

Due to the characteristics of quantum mechanics, the state of a quantum program cannot be directly observed without destroying it.
To nevertheless check the correctness of a quantum program at runtime, several techniques have been proposed.
Figure~\ref{fig:assertion_circuits_for_entangled_states} shows an example circuit (top) with three assertion-checking approaches (bottom): (a) statistical assertions, (b) ancilla-based assertions, and (c) projective assertions.
The following presents the three approaches.

\subsubsection{\statAssertion{} Assertions}
Proposed by \citet{huangStatisticalAssertionsValidating2019}, the idea of statistical assertions is to run the program up to a breakpoint and then measure all the qubits in the program.
Because measuring destroys the quantum state, the remaining program execution is ignored.
This process is repeated multiple times and the output distribution is compared against a specific reference distribution.
This method supports assertions on three kinds of quantum states: classical states, superposition states, and entangled states.
Examples of these states are shown in Figure~\ref{fig:assertion_circuits_for_entangled_states} (top) as $\ket{\psi_1}$, $\ket{\psi_2}$, and $\ket{\psi_3}$, respectively.
In $\ket{\psi_1} = \ket{00}$, the reference distribution is a classical distribution, i.e., all the probability mass is concentrated on a single outcome.
In $\ket{\psi_2} = \frac{1}{\sqrt{2}} \ket{00} + \frac{1}{\sqrt{2}} \ket{10}$, the reference distribution is a uniform distribution across two possible outcomes, $\ket{00}$ and $\ket{10}$, while the probability of any other outcome is zero.
In $\ket{\psi_3} = \frac{1}{\sqrt{2}} \ket{00} + \frac{1}{\sqrt{2}} \ket{11}$, the reference distribution is a uniform distribution across two possible outcomes, $\ket{00}$ and $\ket{11}$, and the probability of any other outcome is zero.
Note that in $\ket{\psi_3}$ the two qubits are entangled because whenever the first qubit is $\ket{0}$ the second qubit is also $\ket{0}$, and likewise for $\ket{1}$.

\subsubsection{\ancillaAssertion{} Assertions}
Inspired by quantum error correction, usually implemented in hardware, \citet{liuQuantumCircuitsDynamic2020} propose \emph{assertion circuits}.
This circuit uses ancilla qubits and indirect measurements to assert the same properties as \citet{huangStatisticalAssertionsValidating2019}, but without stopping the program execution at the breakpoint.
Figure~\ref{fig:assertion_circuits_for_entangled_states}b shows an example of an assertion circuit.
Measuring the ancilla qubits does not disturb the quantum state of the SUT's qubits.
Thus, the state $\ket{\psi_3} = \frac{1}{\sqrt{2}} \ket{00} + \frac{1}{\sqrt{2}} \ket{11}$ is preserved as the output of the assertion circuit.
The only effect of the assertion circuit is that the ancilla qubits are now in a specific state that depends on the outcome of the measurement.
Specifically, the ancilla qubit, initialized as $\ket{0}$, remains $\ket{0}$ if the assertion holds; otherwise, it flips to $\ket{1}$.
Unlike statistical assertions, assertion circuits can check a qubit's phase, which is not observable with a simple output distribution.
However, they can only check assertions where all states in the entanglement have the same parity, i.e., the number of 1s in the bit string is either all even or all odd.
For instance, the entangled state $\ket{\psi} = a \ket{000} + b \ket{011} + c \ket{101} + d \ket{111}$, where $a$, $b$, $c$, and $d$ are complex numbers, cannot be asserted since $\ket{111}$ has different parity.

\subsubsection{\projectiveAssertion{} Assertions}
\citet{liProjectionbasedRuntimeAssertions2020} propose Proq and \emph{projective measurements}, or short \emph{projections}.
Their method ensures that if the quantum state is \ket{phi_a} and we apply the projective measurement $P_A$, which can be viewed as an assertion, then the state \ket{phi_a} remains unchanged.
Since quantum computers typically implement only projections in the computational basis, i.e., those that return either a state \ket{0} or \ket{1} for a single qubit, Proq introduces an approach called \emph{implementation in the computational basis} to allow for more general assertions.
The key idea is to add unitary transformations before and after the projective measurement to implement arbitrary projective measurements, as shown in Figure~\ref{fig:assertion_circuits_for_entangled_states}c.
This approach can also assert new types of quantum states, such as general entangled states, not supported by assertion circuits. For highly entangled states with many qubits, where the unitary implementation is tricky to obtain, they introduce the approximation of \emph{local projections}, checking only a subset of qubits at a time.
This approach is close in spirit to the approach based on combinatorial testing proposed by \citet{wangApplicationCombinatorialTesting2021}, where only pairs or triplets of qubits are considered at the same time.

Finally, building on \citet{liuQuantumCircuitsDynamic2020} and \emph{non-destructive discrimination} (NDD) of quantum states~\cite{jainSecureQuantumConversation2009, guptaGeneralCircuitsIndirecting2007}, \citet{liuSystematicApproachesPrecise2021} propose swap-based and NDD-based methods to check quantum program correctness via ancilla measurements. These methods also support new assertion types, such as mixed states and membership assertions.
Indeed, besides the vector representation \ket{\psi} of a quantum state (Section~\ref{sec:background}), there is also the \emph{density matrix} $\rho$, obtained via the outer product of the state with itself: $\rho = \ket{\psi}\bra{\psi}$.
A state is \emph{pure} if it can be described with a single vector, whereas a state is \emph{mixed} if it is a weighted sum of pure states: $\rho = \sum_i p_i \ket{\psi_i}\bra{\psi_i}$ where $p_i$ is the probability of the state $\ket{\psi_i}$.
Note that a pure state is a special case of a mixed state where $p_i = 1$ for a single $i$.
Mixed states often model noise in quantum programs.
The mixed state assertion checks if a quantum state is a mixture of pure states, whereas the membership assertion checks if a quantum state is a member of a set of quantum states.

\newcommand{\colwidth}{1.2cm}
\begin{table}[t]
  \caption{Types of assertions supported by different approaches: fully (\checkmark), partially ($\approx$), or not supported (empty). Assertions can be on classical, superposition, entangled, or membership states. Limitations: (a) state destruction, (b) mid-circuit measurement, (c) extra qubits, (d) extra operations.}
  \centering
  \begin{tabular}{lp{\colwidth{}}p{\colwidth{}}p{\colwidth{}}p{\colwidth{}}cccc}
  \toprule
  & \multicolumn{4}{c}{\textbf{Types of Assertions}} & \multicolumn{4}{c}{\textbf{Limitations}} \\
  \cmidrule(lr){2-5} \cmidrule(lr){6-9}
  \textbf{Approach}  & \makecell{\textbf{Class.}} & \makecell{\textbf{Superp.}} & \makecell{\textbf{Entang.}} & \makecell{\textbf{Memb.}} & \makecell{\textbf{State} \\ \textbf{destr.}} & \makecell{\textbf{Mid-circ.} \\ \textbf{Meas.}} & \makecell{\textbf{Extra} \\ \textbf{qub.}} & \makecell{\textbf{Extra} \\ \textbf{ops.}} \\
  \midrule
  \multicolumn{9}{l}{\textit{\statAssertion{} assertions:}} \\
  Stat.Assert.~\cite{huangStatisticalAssertionsValidating2019} & \checkmark & $\approx$ & $\approx$ &  & \checkmark &  &  &  \\
  \midrule
  \multicolumn{9}{l}{\textit{\ancillaAssertion{} assertions:}} \\
  Runtime Assertion~\cite{liuQuantumCircuitsDynamic2020} & \checkmark & \checkmark & $\approx$ &  &  &  & \checkmark & \checkmark \\
  Swap-Based~\cite{liuSystematicApproachesPrecise2021} & \checkmark & \checkmark & \checkmark & $\approx$ &  &  & \checkmark & \checkmark \\
  NDD-Based~\cite{liuSystematicApproachesPrecise2021} & \checkmark & \checkmark & \checkmark & $\approx$ &  &  & \checkmark & \checkmark \\
  \midrule
  \multicolumn{9}{l}{\textit{\projectiveAssertion{} assertions:.}} \\
  Proq~\cite{liProjectionbasedRuntimeAssertions2020} & \checkmark & \checkmark & \checkmark &  &  & \checkmark &  & \checkmark \\
  \bottomrule
  \end{tabular}
  \label{tab:assertion_types}
  \end{table}

The approaches discussed above support different types of assertions and have different limitations, as summarized in Table~\ref{tab:assertion_types}.
In terms of scalability, \emph{statistical assertions} require measurements proportional to the number of assertion points, as the program cannot proceed further.
In contrast, methods based on \emph{assertion circuits} and \emph{projective measurements} do not increase the number of required measurements, since runtime assertions leave the state intact.

\subsection{End-of-Computation Oracles}
\label{sec:testing_end_of_computation}

An end-of-computation oracle checks the correctness of the output of a program at the end of the computation.
We discuss such oracles for testing both quantum programs and quantum computing platforms.
For quantum programs, these oracles differ from in-computation assertions in two ways.
First, the oracle can measure and destroy the quantum state, as no further computation follows.
Second, end-of-computation oracles typically check the entire output distribution, not just part of the quantum state.
For quantum platforms, an end-of-computation oracle usually checks the result of compiling or transforming a quantum program.

End-of-computation oracles relate to two terms from physics: quantum property testing and quantum state tomography.
\emph{Quantum property testing}~\cite{montanaroSurveyQuantumProperty2016} checks if a quantum state is close to an expected state, focusing on efficiency and theoretical contexts, like verifying a quantum algorithm properties.
In contrast, end-of-computation oracles apply to concrete quantum algorithm implementations.
\emph{Quantum state tomography}~\cite{nielsenQuantumComputationQuantum2011} reconstructs a quantum state's representation through multiple measurements on identical system copies, requiring many program runs.
While quantum state tomography can serve as an end-of-computation oracle, it is costly, as measurements grow exponentially with qubits.

\begin{figure}
  \begin{tikzpicture}[
    edge from parent/.style = {draw, -latex},
    level distance = 1cm,
    sibling distance = 4.8cm
    ]

  \node {End-of-computation oracles}
    child { node {Precise oracles~\cite{aliAssessingEffectivenessInput2021,paltenghiMorphQMetamorphicTesting2023,abreuMetamorphicTestingOracle2022}} }
    child { node {Imprecise oracles}
      child { node {Distance measures~\cite{dingSQUAREStrategicQuantum2020,patelQUESTSystematicallyApproximating2022,muqeetMitigatingNoiseQuantum2024,wangQDiffDifferentialTesting2021,nishioExtractingSuccessIBM2020}} }
      child { node {Statistical tests~\cite{aliAssessingEffectivenessInput2021,wangQDiffDifferentialTesting2021,wangApplicationCombinatorialTesting2021, wangGeneratingFailingTest2021}} }
    child { node {Equivalence oracles~\cite{pehamEquivalenceCheckingQuantum2022,xuQuartzSuperoptimizationQuantum2022,longEquivalenceIdentityUnitarity2024,unruhQuantumRelationalHoare2019,patelQUESTSystematicallyApproximating2022}} }
  };
  \end{tikzpicture}
  \caption{Overview of approaches that use or propose end-of-computation oracles.}
  \label{fig:oracles summary}
\end{figure}
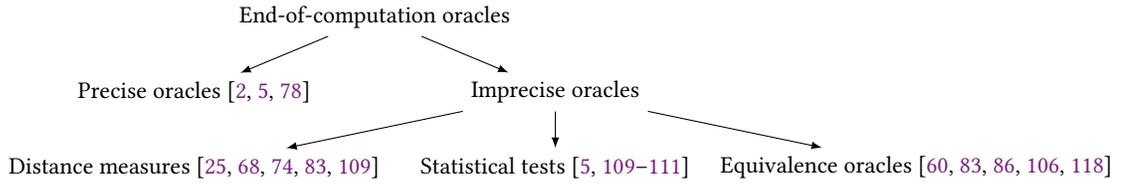

The following discusses approaches to implement end-of-computation oracles in three groups.
Section~\ref{sec:precise oracles} presents \emph{precise oracles}, which reliably identify specific failing cases, detecting bugs in the SUT.
Section~\ref{sec:imprecise oracles} presents \emph{imprecise oracles}, which characterize the correctness of the entire output distribution.
We recognize that this distinction can blur, as some methods serve both purposes.
However, we find it useful to distinguish them due to their different goals and requirements.
Finally, Section~\ref{sec:equivalence oracles} discusses \emph{equivalence oracles}, used to check if two quantum programs are equivalent.
Figure~\ref{fig:oracles summary} provides an overview of the discussed approaches.

\subsubsection{Precise Oracles}
\label{sec:precise oracles}
The precise oracles are designed to find specific bugs in the SUT by observing clear indicators of failure.
For instance, \citet{aliAssessingEffectivenessInput2021} propose the wrong output oracle, which flags a bug every time the program produces an output that is impossible under the correct expected program execution, such as an output with zero probability in the program specification.
This approach assumes a noise-free environment, as NISQ architectures may occasionally produce incorrect output bit strings.
\citet{abreuMetamorphicTestingOracle2022} apply the idea of metamorphic testing~\cite{chenMetamorphicTestingNew2020} to quantum programs, where a metamorphic oracle is a function that takes the output of the SUT and the input and returns a boolean value indicating if the output is correct.
\citet{paltenghiMorphQMetamorphicTesting2023} focuses on a traditional crash oracle, where a bug in the quantum computing platform is signaled by a crash during the execution of a transpiled quantum program.

\subsubsection{Imprecise Oracles}
\label{sec:imprecise oracles}
Imprecise oracles assess the overall correctness of the output distribution. They use two main strategies: (i) distance measures and (ii) statistical tests. Next we present examples and discuss shared challenges.

\paragraph*{Distance Measures}
To compare the output distribution of a quantum program with the expected distribution, several distance measures have been proposed.
Formally, the task is to compare two probability distributions $P$ and $Q$ to determine their similarity.
Each distribution is a vector of probabilities, where the $i$-th element represents the probability of the $i$-th outcome, i.e., the probability of observing the $i$-th bit string as output.
For a program with $n$ qubits, there are $2^n$ possible outcomes.
Among the distance measures are the Hellinger distance, cross-entropy, Kullback-Leibler divergence, Jensen-Shannon divergence, and Total Variation distance. Notably, not all are symmetric, e.g., Kullback-Leibler is asymmetric, while Jensen-Shannon is symmetric, combining two Kullback-Leibler divergences with the average distribution $M = \frac{1}{2} (P + Q)$.
Fidelity, instead of using probability distributions directly, works with the density matrices $\rho$ and $\sigma$ of the quantum states corresponding to $P$ and $Q$.
Thus, we first estimate the density matrices from the distributions using methods like quantum state tomography.

\paragraph*{Statistical Test}
Common statistical tests include the Kolmogorov-Smirnov test, Pearson's chi-squared test, and Wilcoxon signed-rank test.
The Kolmogorov-Smirnov test is non-parametric for continuous distributions, using cumulative distribution functions $\mathcal{P}$ and $\mathcal{Q}$ of $P$ and $Q$.
Here, $\mathcal{P}_i$ is the fraction of samples in $P$ less than or equal to bit string $i$, with bit strings sorted by their decimal values, e.g., $000 \rightarrow 0$, $001 \rightarrow 1$, $010 \rightarrow 2$.
While converting bit strings to decimal values is necessary, this sorting may not be optimal, as bit strings are not independent.
Alternative versions for discrete~\cite{dimitrovaComputingKolmogorovSmirnovDistribution2020} and multivariate~\cite{justelMultivariateKolmogorovSmirnovTest1997} cases exist.
For quantum state verification, \citet{yuStatisticalMethodsQuantum2022} discuss in detail which statistical methods can verify the correctness of a quantum program's output distribution.

\paragraph*{Thresholds and Sample Sizes}
Both statistical tests and distance measures require a threshold to determine if output distributions differ significantly, indicating a bug.
Distance measures use a heuristic threshold to assess deviation from the expected output.
Statistical tests yield a p-value, showing the likelihood of an observed result under the null hypothesis (i.e., the program is correct). A p-value below 0.05 typically indicates a significant difference in output distribution.
Estimating the right number of samples is still an open problem, but \citet{wangQDiffDifferentialTesting2021} propose a formula for the Kolmogorov-Smirnov test, based on the fact that Kolmogorov-Smirnov distance is bounded by L1 distance.

\subsubsection{Equivalence Oracles}
\label{sec:equivalence oracles}
When the expected output can be described by a \emph{reference program}, we can use \emph{equivalence checking} oracles to verify if the SUT's computation matches the reference.
\citet{pehamEquivalenceCheckingQuantum2022} uses ZX-calculus to check the equivalence of two quantum programs for quantum compilation.
\citet{xuQuartzSuperoptimizationQuantum2022} proposes an approach based on SAT solver to discover new equivalences between chains of gates.
In their work, \citet{longEquivalenceIdentityUnitarity2024} propose algorithms to check program equivalence, assess if a program is unitary, namely reversible, and to determine whether a program is the identity.
\citet{unruhQuantumRelationalHoare2019} proposes a relational version of Hoare logic (rQHL) for quantum programs, which can be used to prove the equivalence of two programs.
A technique called Feynman~\cite{amyLargescaleFunctionalVerification2019} allows to check the equivalence of two Clifford group quantum programs in polynomial time, by using a simulation approach based on the path sum representation.
It also handles circuits of this kind with up to 100 qubits and thousands of gates.
Quest~\cite{patelQUESTSystematicallyApproximating2022} verifies approximate semantic equivalence of two programs using the Hilbert-Schmidt distance between their unitary matrices.

\subsection{Debugging of Quantum Programs}
\label{sec:debugging}
Debugging, i.e., identifying and understanding bugs in a program, is crucial in software development. For quantum programs, it is particularly challenging due to the probabilistic nature of quantum mechanics and the no-cloning theorem, which prevents copying an unknown quantum state. Yet, there is relatively little work to support debugging of quantum programs.
\citet{metwalliToolDebuggingQuantum2022} propose to automate the task of running only parts of a quantum program, letting users exclude qubits and select circuit segments to execute.
Another method~\cite{suauQprofGprofInspiredQuantum2022} profiles quantum programs similarly to traditional CPU profilers.
Their method informs the user how often a subroutine is called, highlighting optimization opportunities like removing unused subroutines or replacing them with more efficient ones.

\section{Static Analysis for Correctness}
\label{sec:static_analysis_correct}
This section covers static analysis methods to detect bugs or prove correctness without executing the software on a quantum computer, unlike the previous section on testing techniques requiring execution on a quantum device.
We also discuss approaches that run quantum programs on simulators, considered a form of static analysis since they do not use real quantum computers.

Table~\ref{tab:families_correctness_vs_sut} summarizes the discussed approaches, categorized by analyzed software (quantum programs or platforms) and analysis technique.
We identify three kinds of analysis techniques:
First, \emph{proof-based formal methods}, which aim at exact reasoning about computations, e.g., based on Hoare logic, separation logic, and incorrectness logic, and rely on theorem provers or deductive frameworks.
Second, \emph{formal methods based on abstraction}, such as abstract interpretation, that may rely on higher degrees of approximation of the quantum state than the previous group.
Third, \emph{pattern-based approaches} that look for recurring bug patterns, e.g., in the form of source code fragments that have a specific syntactic structure, specific data flow relations between code elements, or a specific sequence of quantum gates.
These approaches trade off precision and scalability: proof-based methods are most precise but least scalable, while pattern-based methods are more scalable but less precise.

Several approaches adapt classical static analysis techniques to quantum, including quantum Hoare logic~\cite{yingFloydHoareLogic2012}, abstract interpretation~\cite{yuQuantumAbstractInterpretation2021}, symbolic execution~\cite{nanQuantumSymbolicExecution2023}, syntax-based analysis~\cite{zhaoQCheckerDetectingBugs2023a}, and data flow-based analysis~\cite{paltenghiAnalyzingQuantumPrograms2024,kaulUniformRepresentationClassical2023a}.
Other analyses are specifically developed for quantum programs, such as \emph{entanglement analysis}~\cite{perdrixQuantumEntanglementAnalysis2008, yuanTwistSoundReasoning2022, xiaStaticEntanglementAnalysis2023}, which is used to track which qubits are entangled with each other, and \emph{uncomputation analysis}~\cite{dingSQUAREStrategicQuantum2020, paradisUnqompSynthesizingUncomputation2021}, which tracks qubits that are used as ancilla and are still entangled with the rest of the program state, and thus cannot be used for other purposes.

Sections~\ref{sec:formal_methods_proof_based}, \ref{sec:formal_methods_abstraction}, and \ref{sec:pattern_based} discuss these three families of approaches.
We cover their representation of quantum computation semantics, the properties they check, and their automation levels.

\begin{table}[t]
    \centering
    \caption{Overview of analysis methodologies and target System Under Test (SUT).}
    \label{tab:families_correctness_vs_sut}
    \begin{tabular}{p{1.5cm}p{4cm}p{4cm}p{4cm}}
        \toprule
                     & \multicolumn{3}{c}{\textbf{Kinds of analysis approaches}}                                                                                                                                                                                                                                            \\
        \cmidrule{2-4}
        \textbf{Analyzed software} & \textbf{Proof-based formal methods (Section~\ref{sec:formal_methods_proof_based})}                                                                             & \textbf{Formal methods based on abstraction (Section~\ref{sec:formal_methods_abstraction})} & \textbf{Pattern-based approaches (Section~\ref{sec:pattern_based})} \\
        \midrule
        Program      &
        Incorrectness logic \cite{yanIncorrectnessLogicQuantum2022}, Quantum Hoare logic \cite{yingFloydHoareLogic2012, zhouAppliedQuantumHoare2019},
        QHLProver~\cite{liuFormalVerificationQuantum2019}, Interactive provers~\cite{hietalaProvingQuantumPrograms2021,bordgCertifiedQuantumComputation2021}, QWIRE~\cite{randQWIREPracticeFormal2018}, Qbricks~\cite{charetonAutomatedDeductiveVerification2021},
        CoqQ~\cite{zhouCoqQFoundationalVerification2023}, SymQV~\cite{bauer-marquartSymQVAutomatedSymbolic2023},
        Quantum separation logic \cite{zhouQuantumInterpretationBunched2021}
                     & Quantum abstract interpretation \cite{yuQuantumAbstractInterpretation2021},
        Abstraqt~\cite{bichselAbstraqtAnalysisQuantum2023},
        AutoQ~\cite{chenAutomataBasedFrameworkVerification2023,chenAutoQAutomataBasedQuantum2023},
        Entanglement analysis~\cite{perdrixQuantumEntanglementAnalysis2008,xiaStaticEntanglementAnalysis2023}, Twist~\cite{yuanTwistSoundReasoning2022}, ScaffCC~\cite{javadiabhariScaffCCFrameworkCompilation2014}
                     & Qchecker~\cite{zhaoQCheckerDetectingBugs2023a}, LintQ~\cite{paltenghiAnalyzingQuantumPrograms2024}, Quantum-CPG~\cite{kaulUniformRepresentationClassical2023a},
        QSmell~\cite{chenSmellyEightEmpirical2023}                                                                                                                                                                                                         \\

        Platform     &
        VOQC~\cite{hietalaVerifiedOptimizerQuantum2021,hietalaVerifiedOptimizerQuantum2023},
        Giallar~\cite{taoGiallarPushbuttonVerification2022}, ReQwire~\cite{randReQWIREReasoningReversible2019}
                     & -
                     & -
        \\
        \bottomrule
    \end{tabular}
\end{table}

\subsection{Proof-Based Formal Methods}
\label{sec:formal_methods_proof_based}

The goal of the approaches discussed in the following is to formally specify properties of quantum programs and prove these properties using formal methods, such as logic, theorem provers, or SMT solvers.

\subsubsection{Representations of Computations}
\label{sec:proof_based_representation}

\begin{figure}[t]
    \centering
    \includegraphics[width=\textwidth]{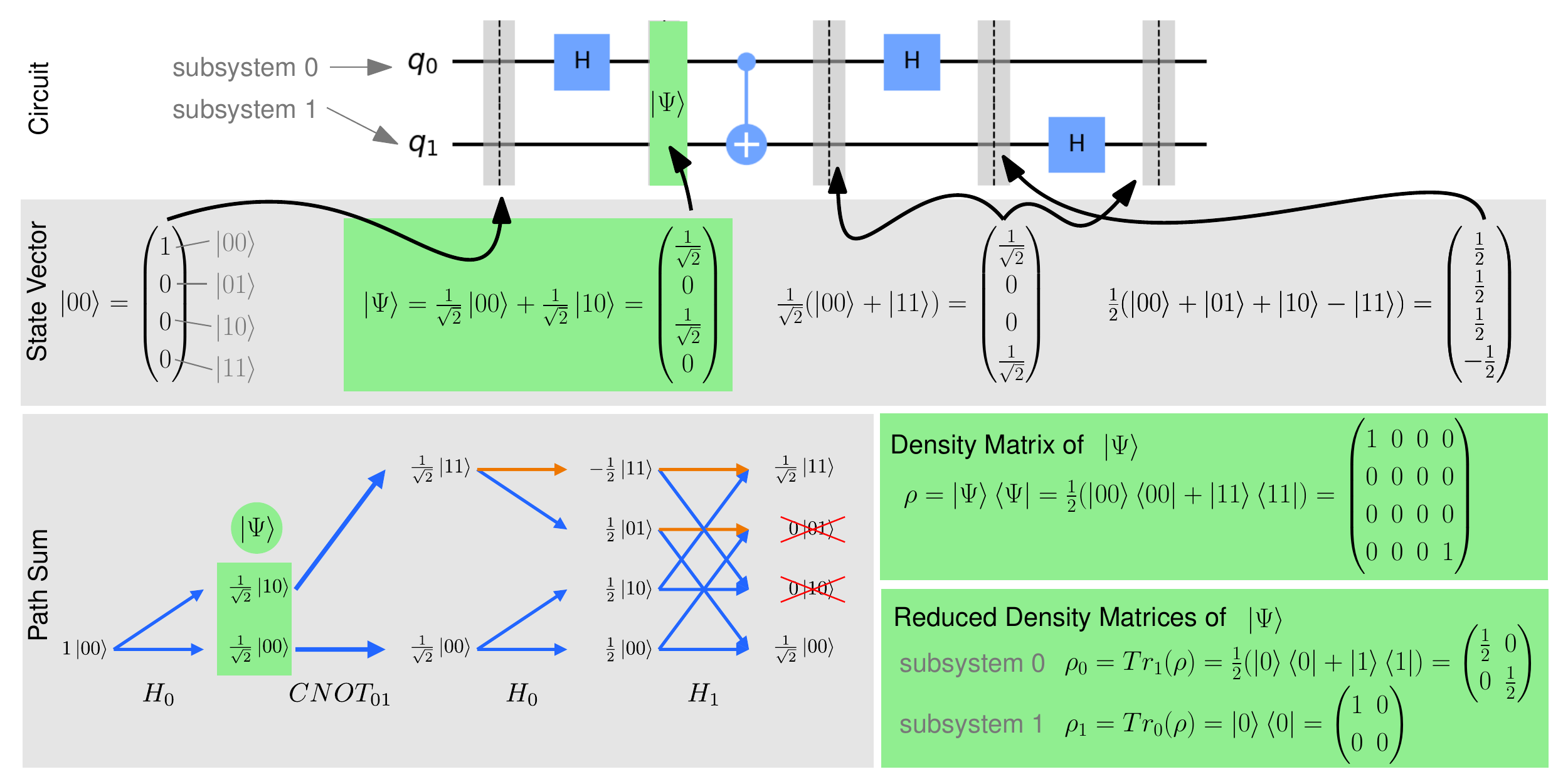}
    \caption{Quantum circuit and various representations of quantum states during computation. The state vector and path sum represent the entire circuit, while the green boxes show a specific state $\ket{\psi}$ with a (reduced) density matrix.}
    \label{fig:sa_correct_ch_overview_representation}
\end{figure}

Different approaches use various representations for quantum states and operations, including state vectors, path sums, and (reduced) density matrices.
Figure~\ref{fig:sa_correct_ch_overview_representation} shows common representations with an example.
We first explain these representations and then discuss their use in existing techniques.

The top of Figure~\ref{fig:sa_correct_ch_overview_representation} shows the quantum circuit, with different quantum state representations below.
The upper gray box in Figure~\ref{fig:sa_correct_ch_overview_representation} shows the state vector representation at different points in the circuit.
A state vector represents the quantum state as a vector $\ket{\Psi}$ of complex amplitudes, one for each basis state.
The lower gray box shows the path sum representation, which expresses a quantum state as a sum of all possible paths leading to it.
Each path is a sequence of possible quantum states from the initial to the final states.
Finally, the two green boxes at the bottom right show the density matrix of a specific state, marked in green throughout the figure.
The top one shows the full density matrix $\rho = \ket{\Psi}\bra{\Psi}$, while the bottom one shows the reduced density matrix.

State vectors and density matrices are widely used due to their accuracy~\cite{yanIncorrectnessLogicQuantum2022, yingFloydHoareLogic2012, liuFormalVerificationQuantum2019, zhouAppliedQuantumHoare2019, hietalaProvingQuantumPrograms2021}.
The downside of these representations is their computation and memory cost which is exponential in the number of qubits.
\citet{bordgCertifiedQuantumComputation2021} use a state vector modeled as complex numbers in the Isabelle theorem prover, but it has limited power since it cannot represent mixed states. CoqQ~\cite{zhouCoqQFoundationalVerification2023} uses the more expressive Dirac notation, which is easier for reasoning when using the MathComp library.
To avoid the exponential cost of state vectors and density matrices, other approaches, like the Feynman tool~\cite{amyLargescaleFunctionalVerification2019} and Qbricks~\cite{charetonAutomatedDeductiveVerification2021}, use path sums. Another method, SymQV~\cite{bauer-marquartSymQVAutomatedSymbolic2023}, employs an SMT-compatible symbolic state representation.
Here, the state is represented as a set of symbolic variables that represent the qubits of the programs and that are manipulated by the quantum gates.
This representation reduces computational cost and memory usage by avoiding full state vector and matrix construction when possible.

\subsubsection{Kinds of Checked Properties}
\label{sec:proof_based_properties}
The properties checked by different approaches vary significantly, from verifying the correctness of a specific QFT implementation~\cite{amyLargescaleFunctionalVerification2019} to fundamental quantum computing theorems like the no-cloning theorem, which states that an exact copy of an unknown quantum state cannot be made~\cite{bordgCertifiedQuantumComputation2021}.

Inspired by classical Hoare logic~\cite{hoareAxiomaticBasisComputer1969}, \citet{yingFloydHoareLogic2012} formalizes quantum Hoare logic (QHL).
Like its classical counterpart, a quantum Hoare triple is $\{P\}C\{Q\}$, where $P$ and $Q$ are quantum predicates for pre- and post-conditions, and $C$ is the quantum program.
Several approaches build on QHL~\cite{zhouAppliedQuantumHoare2019, liuFormalVerificationQuantum2019,yingFloydHoareLogic2012}, differing in the supported predicates $P$ and $Q$, e.g., arbitrary quantum predicates~\cite{yingFloydHoareLogic2012, liuFormalVerificationQuantum2019} or projections~\cite{zhouAppliedQuantumHoare2019}.
Projections are specific quantum predicates, defined as $P = \sum_{i=1}^{n} \ket{i}\bra{i}$, where $n$ is the number of qubits and $\ket{i}$ is the computational basis.
Projections enable simpler inference rules, and hence, smaller proof trees, but cannot capture all quantum properties.
\citet{zhouAppliedQuantumHoare2019}, in their work on \emph{applied} quantum Hoare logic (aQHL) use projections to simplify inference rules and ranking functions of QHL.
They also introduce rules for reasoning about robustness, useful for inexact quantum programs like those in quantum machine learning.
Most approaches allow expressing first-order logic properties of quantum programs, making them versatile for proving properties.
Two examples of provable correctness properties are:
(1) Grover's algorithm has a higher probability of measuring the correct solution than a threshold, given a fixed number of iterations~\cite{hietalaProvingQuantumPrograms2021};
(2) the HLL algorithm returns the exact solution for any input state, given certain assumptions about the solution's existence~\cite{zhouAppliedQuantumHoare2019}.

A dual view of Hoare logic is \emph{incorrectness logic}~\cite{ohearnIncorrectnessLogic2019}, which aims to prove the existence of bugs rather than program correctness.
It is based on the triple $\{presumption\}C\{result\}$, where the presumption is a predicate assumed true before execution, and the result is a predicate true on a subset of possible final states.
Inspired by incorrectness logic, \citet{yanIncorrectnessLogicQuantum2022} propose \emph{quantum incorrectness logic} (QIL), using projections for presumption and result.
Another logic adapted to quantum is separation logic~\cite{reynoldsSeparationLogicLogic2002}, which reasons about program memory modularly.
Separation logic has been adapted to quantum to reason about entanglement and separability of subsystems~\cite{zhouQuantumInterpretationBunched2021}.

The Feynman tool~\cite{amyLargescaleFunctionalVerification2019} allows proving the equivalence of two quantum programs or that two quantum programs implement different functions.
Qbrick~\cite{charetonAutomatedDeductiveVerification2021}, extending Feynman's state representation, allows expressing first-order logic properties.
Both Qbrick~\cite{charetonAutomatedDeductiveVerification2021} and QWIRE~\cite{randQWIREPracticeFormal2018} allow for specifying properties of parametric quantum programs, e.g., an single algorithm that can operate on a variable number of qubits, using two different theorem provers, namely Why3 and Coq.
Finally, SymQV~\cite{bauer-marquartSymQVAutomatedSymbolic2023} can verify first-order logic properties of quantum programs.
For instance, it can check that in the quantum teleportation protocol, the state of the first qubit is transferred to the last qubit, and no operations cross the boundary between the first two qubits and the last one.
Namely, the first two qubits are the sender's, and the last one is the receiver's.

\subsubsection{Level of Automation}
\label{sec:proof_based_automation}

The above techniques rely on axiomatic proof systems, with automation levels ranging from manual to fully automatic. Early methods are mostly manual or semi-automatic, requiring users to construct proofs and provide intermediate lemmas.
This applies to Hoare logic~\cite{yingFloydHoareLogic2012}, incorrectness logic~\cite{ohearnIncorrectnessLogic2019}, and SQIR~\cite{hietalaProvingQuantumPrograms2021}, in which verifying Grover and QPE took one and two person-weeks, respectively.
Qbricks~\cite{charetonAutomatedDeductiveVerification2021} automates proofs using a domain-specific language (Qbricks-DSL) for quantum programs and another for specifications (Qbricks-SPEC). It automatically proves 95\% of lemmas and proof obligations in their evaluation. %
Two fully automated approaches are Giallar~\cite{taoGiallarPushbuttonVerification2022} and SymQV~\cite{bauer-marquartSymQVAutomatedSymbolic2023}.
Giallar uses SMT solvers to verify circuit transformations by an optimizer in a quantum platform.
It extracts invariants from three loop templates and uses pre-verified rewriting rules to prove circuit transformation correctness.
SymQV~\cite{bauer-marquartSymQVAutomatedSymbolic2023} is a symbolic execution framework that proves first-order logic properties of quantum programs using an SMT solver.
A potential barrier to adoption is that many approaches require users to use a specific programming language, often custom or domain-specific, such as the ``quantum while language''~\cite{yingFloydHoareLogic2012} or Qbricks-DSL~\cite{charetonAutomatedDeductiveVerification2021}, or intermediate representations embedded in theorem provers like SQIR~\cite{hietalaVerifiedOptimizerQuantum2021,hietalaVerifiedOptimizerQuantum2023} and QWIRE~\cite{randQWIREPracticeFormal2018}.
Most languages are low-level and tailored to specific methods, except SQIR, which converts to and from OpenQASM.

\subsection{Formal Methods based on Abstraction}
\label{sec:formal_methods_abstraction}

The following discusses approaches that use formal methods but rely on abstractions to reduce analysis complexity. These methods represent the quantum state abstractly, simplifying reasoning but potentially losing some information.

\subsubsection{Representations of Computations}
\label{sec:abstraction_based_representation}
A common analysis target here is entanglement, tracking which qubits are entangled. Due to the exponential memory needed for precise state tracking, the \emph{stabilizer} formalism~\cite{aaronsonImprovedSimulationStabilizer2004} is often used.
This formalism represents the quantum state $\ket{\psi}$ using Pauli operators $Q$ that stabilize it, satisfying $Q \ket{\psi} = \ket{\psi}$.
While compact and efficient, it is limited to states produced by Clifford gates, which are not universal for quantum computation.
To handle more programs, \citet{perdrixQuantumEntanglementAnalysis2008} use diagonal bases to abstract the quantum state, though this is quite lossy. Inspired by stabilizers' precision, \citet{hondaAnalysisQuantumEntanglement2015} extend stabilizers to allow some uncertainty when non-Clifford gates like the T gate are used.
The key idea is that non-Clifford gates' effects can be bounded locally and removed later. While \citet{hondaAnalysisQuantumEntanglement2015} is more precise than \citet{perdrixQuantumEntanglementAnalysis2008}, its use beyond entanglement analysis remains unclear.
Twist~\cite{yuanTwistSoundReasoning2022} instead uses the concept of fractional permissions to model entanglement along the computations, and the usual density matrix to represent the quantum state whenever they need to resolve the quantum state to a specific value at runtime.
ScaffCC~\cite{javadiabhariScaffCCFrameworkCompilation2014} tracks qubits that might be entangled, assuming any two-qubit gate creates entanglement.

Other approaches abstract the quantum state more generally, beyond entanglement.
\citet{yuQuantumAbstractInterpretation2021} use reduced density matrices (Figure~\ref{fig:sa_correct_ch_overview_representation}).
Note that with reduced density matrices, reconstructing the full quantum state is generally impossible, making the representation lossy.
\citet{yuQuantumAbstractInterpretation2021} use reduced density matrices to represent entanglement between qubit pairs.
This saves space, representing 50 qubits with $\binom{50}{2} = 1225$ matrices instead of $2^{50}$ complex numbers.
Their approach handles programs with up to 300 qubits, a significant improvement over naive simulation.
Besides pairs of qubits, the approach can consider any number of qubits together, but this generalization creates multiple abstract domains, complicating their selection for a given program.

Abstraqt~\cite{bichselAbstraqtAnalysisQuantum2023} uses a path sum representation, approximating non-Clifford gates with Pauli sums, then condensing them into a single summand for efficiency. This over-approximation enhances time and space efficiency compared to simulators.
However, the approach loses precision quickly when reasoning about programs with Clifford and T gates.
AutoQ~\cite{chenAutomataBasedFrameworkVerification2023} introduces a binary tree to represent the quantum state, where the height of the tree is the number of qubits, each branch corresponds to a computational basis, and each leaf corresponds to the complex amplitude of that state.
A tree automaton is also used to represent a set of states and the operations via gates are operations directly on the automaton.
AutoQ's tree representation~\cite{chenAutomataBasedFrameworkVerification2023} is space- and computation-efficient.
For example, it can encode the output of the Bernstein-Vazirani algorithm with a linearly sized tree by merging the shared branches of the tree.

\subsubsection{Kinds of Checked Properties}
\label{sec:abstraction_based_properties}
The properties checked by different approaches vary significantly, with the most efficient ones checking only simple properties like entanglement absence.
For example, \citet{perdrixQuantumEntanglementAnalysis2008} checks entanglement between qubit subsets, assuming all CNOT gates entangle.
\citet{hondaAnalysisQuantumEntanglement2015} also checks entanglement between qubit subsets but has limited handling of non-Clifford gates.
Twist~\cite{yuanTwistSoundReasoning2022} instead allows reasoning about entanglement by exploiting annotations of ``pure types'', and assertions that express when some qubits are separable from the rest of the program (called ``cast'' assertion) or when two sets of qubits are not entangled (called ``split'' assertion).
ScaffCC~\cite{javadiabhariScaffCCFrameworkCompilation2014} runs a disentanglement check to ensure all ancilla qubits are not entangled with result qubits.
\citet{yuQuantumAbstractInterpretation2021} checks if a target state is in the span of, i.e., subspace generated by, certain vectors, like the GHZ state $span(\ket{000},\ket{111})$.
This is limited as it cannot represent more complex superpositions.
Abstraqt~\cite{bichselAbstraqtAnalysisQuantum2023} checks if qubits are reset to zero at the program's end
AutoQ~\cite{chenAutomataBasedFrameworkVerification2023} checks if the output state is a sub-language of a tree automaton representing the post-condition, allowing versatile properties to be checked as tree automata.
AutoQ focuses on detecting bugs or functional non-equivalence between two quantum programs rather than proving their correctness.

\subsubsection{Level of Automation}
\label{sec:abstraction_based_automation}
The first two works on abstract interpretation~\cite{perdrixQuantumEntanglementAnalysis2008,hondaAnalysisQuantumEntanglement2015} are fully manual, requiring users to apply abstract semantic rules by hand.
Twist~\cite{yuanTwistSoundReasoning2022} needs type and assertion annotations for quantum expressions but is fully automatic, using static analysis and runtime checks via a simulator.
Similarly, ScaffCC~\cite{javadiabhariScaffCCFrameworkCompilation2014} requires users to annotate ancilla qubits in Scaffold programs. Adding new analyses in ScaffCC requires manual implementation in the compiler.
Yu et al.~\cite{yuQuantumAbstractInterpretation2021} is fully automatic, needing users to specify the span or support of the final state or define loop invariants, as in Grover's algorithm.
Abstraqt~\cite{bichselAbstraqtAnalysisQuantum2023} and AutoQ~\cite{chenAutomataBasedFrameworkVerification2023} are fully automatic, with Abstraqt requiring users to specify correctness properties and AutoQ needing user-defined post-conditions for non-equivalence checks.

\subsection{Pattern-Based Approaches}
\label{sec:pattern_based}

The third group of approaches identifies and matches specific syntactic or semantic constructs, known as \emph{patterns}, in the target language.
Table~\ref{tab:pattern_approaches} summarizes these approaches, which recognize patterns at varying abstraction levels, from purely syntactic to more semantic, like data-flow analysis.

\begin{table}[t]
    \centering
    \caption{Overview of pattern-based approaches.}
    \label{tab:pattern_approaches}
    \begin{tabular}{lcccccc}
        \toprule
        Approach                                                   & LintQ \cite{paltenghiAnalyzingQuantumPrograms2024} & Quantum-CPG \cite{kaulUniformRepresentationClassical2023a} & QChecker \cite{zhaoQCheckerDetectingBugs2023a} & Qsmell \cite{chenSmellyEightEmpirical2023} \\
        \midrule
        Syntax matching (AST)                                      & $\checkmark$                                    & $\checkmark$                                             & $\checkmark$                                 & $\checkmark$                                      \\
        Data flow                                                 & $\checkmark$                                    & $\checkmark$                                             &                                                &                                                  \\
        Quantum data flow                                         & $\checkmark$                                    & $\checkmark$                                             &                                                & $\checkmark$                                      \\
        Control flow                                              & $\checkmark$                                    & $\checkmark$                                             &                                                &                                                  \\
        Circuit composition                                       & $\checkmark$                                    &                                                          &                                                &                                                  \\
        Classical-quantum link                                    &                                                 & $\checkmark$                                             &                                                &                                                  \\
        \bottomrule
    \end{tabular}
\end{table}

\subsubsection{Representations of Computations}
\label{sec:pattern_representation}
All approaches in this category target quantum programs written in Qiskit~\cite{paltenghiAnalyzingQuantumPrograms2024,chenSmellyEightEmpirical2023,zhaoQCheckerDetectingBugs2023a,kaulUniformRepresentationClassical2023a}, thus Python code.
QChecker~\cite{zhaoQCheckerDetectingBugs2023a} analyzes the abstract syntax tree (AST) of quantum programs, representing variable assignments as \emph{QP attributes} and function calls as \emph{QP operations}.
A QP attribute example is a variable assigned to a quantum register, like \code{(`qreg', `QuantumRegister(3)')}; a QP operation example is a function call applied to a quantum circuit, like \code{(`circuit.h(2)')}.
LintQ~\cite{paltenghiAnalyzingQuantumPrograms2024} instead converts the quantum program into the CodeQL intermediate representation and then uses the CodeQL framework to reason about both syntax in the form of the AST, but also derive data-flow and control-flow analysis.
The CodeQL representation consists of a database of facts that encode the program elements and the relationships between them.
Beside classical data flow, LintQ also supports quantum data flow to track gate application order on qubits.'s
Qsmell~\cite{chenSmellyEightEmpirical2023} represents the quantum program using an execution matrix derived from the compiler. In this matrix, rows correspond to qubits, cells to gates, and columns to execution timestamps.
Alternatively, Qsmell supports AST-based analysis without data-flow or control-flow. Quantum-CPG~\cite{kaulUniformRepresentationClassical2023a} converts the quantum program into a code property graph (CPG)~\cite{yamaguchiModelingDiscoveringVulnerabilities2014}, encoding both data-flow and control-flow. Like LintQ, it captures the execution order of quantum gates on qubits using a graph representation of quantum data flow.

\subsubsection{Kinds of Checked Properties}
\label{sec:pattern_properties}
Each approach recognizes different patterns based on the abstraction level and expressiveness of the quantum program representation, as summarized in Table~\ref{tab:bug_patterns}. Despite using different representations, there are overlaps in the patterns they recognize.
Note that even if two approaches recognize the same pattern, they may do it differently, leading to varied bug detection abilities.
QChecker~\cite{zhaoQCheckerDetectingBugs2023a} identifies syntax-level bug patterns using Python's AST, inspired by Bugs4Q dataset~\cite{zhaoBugs4QBenchmarkReal2021}. It focuses on local issues like non-existent APIs or incorrect function parameters.
LintQ~\cite{paltenghiAnalyzingQuantumPrograms2024} recognizes the most patterns, including those from empirical studies and related work.
LintQ is also the only approach that models composed circuits, i.e., complex circuits that are build from  multiple sub-circuits.
Modeling this and other concepts allows LintQ to recognize additional bug patterns, such as code that uses the \code{compose} API incorrectly, and data flow-related patterns, such as applying a gate to a qubit after the qubit has been measured.
QSmell~\cite{chenSmellyEightEmpirical2023} identifies code smells validated by a developer survey. Examples include overly long quantum programs and qubits unused for extended periods, increasing decoherence risk.
Finally, Quantum-CPG~\cite{kaulUniformRepresentationClassical2023a} detects programming errors and classical code smells. It uniquely models the qubit-classical register link, identifying unused measurement results.
We refer to the original papers for the full list of patterns recognized by each approach.

\begin{table}[t]
    \centering
    \caption{Bug patterns recognized by different pattern-based approaches.}
    \label{tab:bug_patterns}
    \begin{tabular}{ll|ll}
        \toprule
        \multicolumn{4}{c}{
            \textbf{Approaches}:
            LintQ (\A) \quad Quantum-CPG (\B) \quad QChecker (\C) \quad QSmell (\D)
        } \\
        \midrule
        \textbf{Bug Pattern} & \textbf{Support} & \textbf{Bug Pattern} & \textbf{Support} \\
        \midrule
        \textbf{Gate errors} & & \textbf{Circuit size} & \\
        Custom / incorrect gates      & \C\D & Classical register too small   & \A\C \\
        Repeated gates                & \D   & Long circuit                   & \D \\
        Superfluous operation         & \B   & Oversized circuit             & \A \\
        \midrule
        \textbf{Measurement issues} & & \textbf{Quantum-classical interface} & \\
        Operation after measurement   & \A\C\D & Constant classical bit        & \A\B \\
        Condition without measurement & \A\B & Non-parametrized circuit      & \D \\
        Double measurement            & \A   & Result bit not used           & \B \\
        Measure all                   & \A   & Constant result bit & \B \\
        \midrule
        \textbf{Miscellaneous} & & \textbf{Qubit initialization and layout} & \\
        Use non-existent API          & \A\C & Idle qubits                    & \D \\
        Parameter error               & \C   & Initialization of qubits       & \D \\
        Call error                    & \C   & No physical layout             & \D \\
        QASM error                    & \C   & & \\
        Ghost compose                 & \A   & & \\
        Operation after optimization  & \A   & & \\
        Discarded order               & \C   & & \\
        \bottomrule
    \end{tabular}
\end{table}

\subsubsection{Level of Automation}
\label{sec:pattern_automation}
Compared to proof-based and abstraction-based methods, pattern-based approaches have the lowest entry barrier. They only need the quantum program's source code and run fully automatically.
Pattern-based approaches can be extended to recognize new patterns with some effort.
Adding patterns to QChecker or QSmell requires Python code to detect patterns in the AST or execution matrix.
For LintQ, it involves writing CodeQL queries using LintQ's abstractions.

\section{Static Analysis to Optimize Quantum Programs}
\label{sec:static_analysis_optimize}

Beyond validating and verifying quantum programs, static analysis can optimize them. Optimization is crucial for NISQ computers~\cite{preskillQuantumComputingNISQ2018}, which face four main challenges:
\begin{itemize}
    \item (C1) \emph{Noisy hardware}. Current gate implementations suffer from a non-negligible error rate. Moreover, values stored in a quantum register suffer from the decoherence effect, where the imperfect isolation of the qubits from their environment may destroy the quantum state, and hence, limit the maximum length of a computation.
    \item (C2) \emph{Constrained resources}. The limited number of qubits demands for a careful allocation of registers.
    \item (C3) \emph{Limited connectivity}. Due to quantum computer's physical design, not all qubits can interact with each other.
    \item (C4) \emph{Gate-set variation}: Since each quantum computer has its own available gates, the same program may appear significantly different when executed on different machines, requiring distinct optimization strategies.
\end{itemize}

\begin{figure}[t]
    \centering
    \caption{Challenges faced when dealing with NISQ computers and corresponding optimization goals pursued by analyses.}
    \label{fig:sa_optimize_ch_overview}
    \begin{subfigure}{0.23\textwidth}
        \centering
        \includegraphics[width=\textwidth]{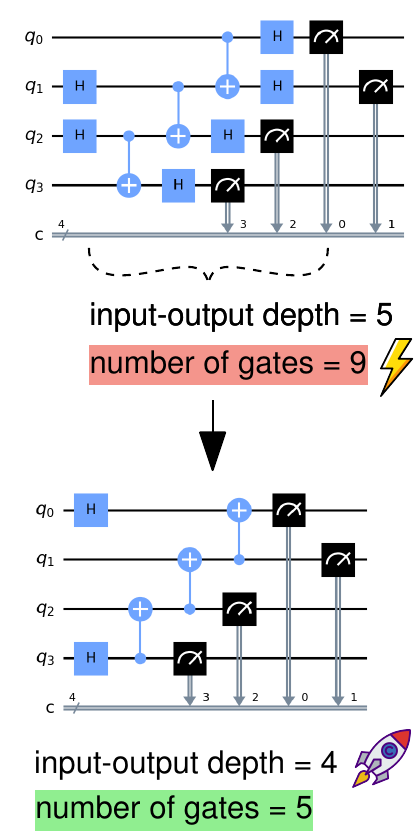}
        \caption{Noisy hardware challenge (C1) and its mitigation by reducing the number of gates (G1).\\}
        \label{fig:sa_optimize_ch_overview_gate_imperfection_and_gate_imperfection}
    \end{subfigure}
    \hspace{0.01\textwidth}
    \begin{subfigure}{0.23\textwidth}
        \centering
        \includegraphics[width=\textwidth]{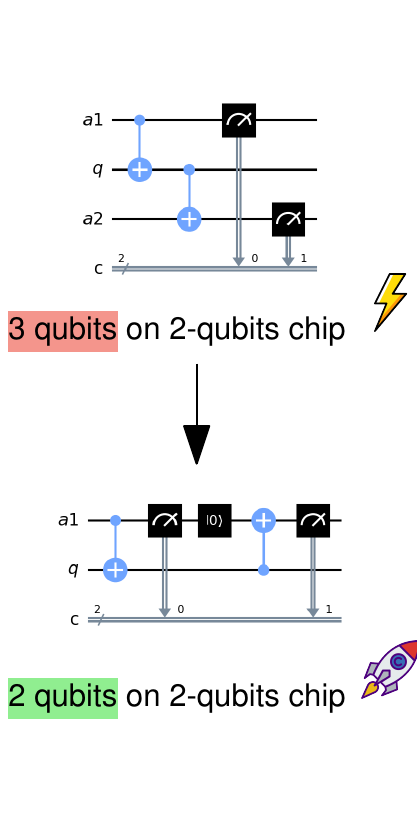}
        \caption{Constrained resources challenge (C2) and its mitigation by reducing the number of qubits (G2).}
        \label{fig:sa_optimize_ch_overview_constrained_resources}
    \end{subfigure}
    \hspace{0.01\textwidth}
    \begin{subfigure}{0.23\textwidth}
        \centering
        \includegraphics[width=\textwidth]{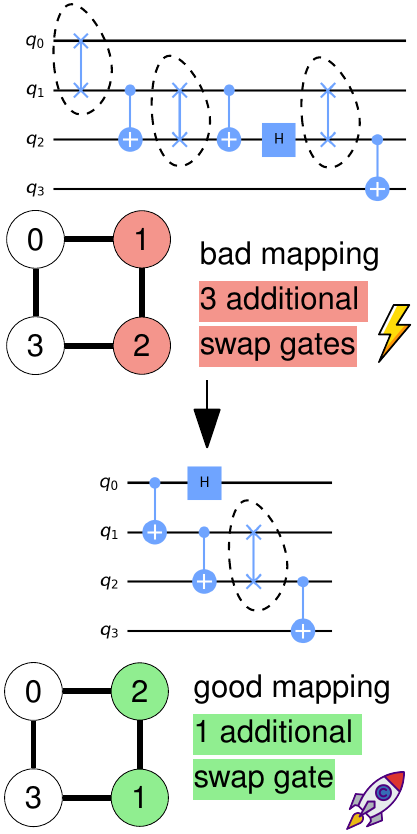}
        \caption{Limited connectivity challenge (C3) and its mitigation by minimizing communication (G3).}
        \label{fig:sa_optimize_ch_overview_limited_connectivity}
    \end{subfigure}
    \hspace{0.01\textwidth}
    \begin{subfigure}{0.23\textwidth}
        \centering
        \includegraphics[width=\textwidth]{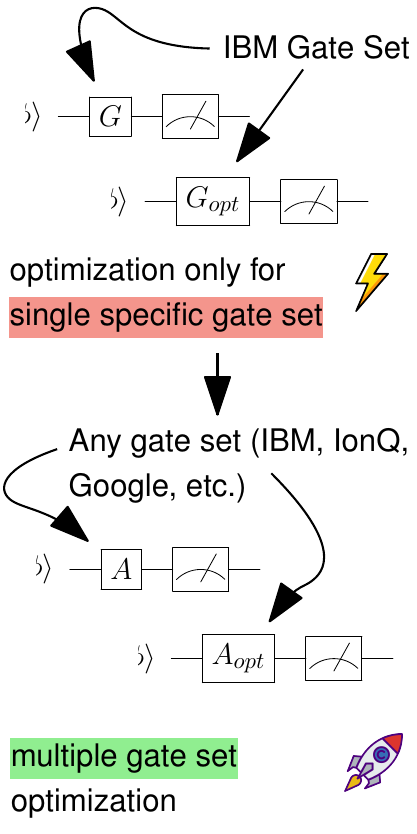}
        \caption{Gate set variation challenge (C4) and its mitigation by being gate set independent (G4).}
        \label{fig:sa_optimize_ch_overview_gate_set_variation}
    \end{subfigure}
\end{figure}

These challenges have inspired various optimization approaches that analyze and transform quantum programs.
We group these approaches by their optimization goals.
As shown in Figure~\ref{fig:sa_optimize_ch_overview}, each challenge (C1--C4) corresponds to a specific optimization goal (G1--G4).
To tackle noisy hardware (C1), optimizations aim to reduce gate count (G1), shortening circuit depth, as shown in Figure~\ref{fig:sa_optimize_ch_overview_gate_imperfection_and_gate_imperfection}.
For constrained resources (C2), methods reduce the number of qubits used (G2), as shown in Figure~\ref{fig:sa_optimize_ch_overview_constrained_resources}.
Addressing limited connectivity (C3) involves optimizing qubit mapping (G3), e.g., by minimizing the use of extra swap gates, as shown in Figure~\ref{fig:sa_optimize_ch_overview_limited_connectivity}.
Finally, to address gate-set variations (C4), optimizations aim for hardware-agnosticism by targeting programs written in any gate set (G4), as shown in Figure~\ref{fig:sa_optimize_ch_overview_gate_set_variation}.

In the following section, we discuss different approaches grouped by their optimization goals, noting that some approaches pursue multiple goals and are discussed in multiple sections. Table~\ref{tab:multiple_optimization_goals} summarizes the optimization goals pursued by different approaches. We exclude error correction and error mitigation work, as they usually do not rely on program analysis but either post-process an algorithm's output or provide an additional hardware mechanism.

\begin{table}[t]
    \centering
    \caption{Approaches for quantum program optimization and their optimization goals}
    \label{tab:multiple_optimization_goals}
    \begin{tabular}{ll|ll}
        \toprule
        \multicolumn{4}{c}{
            \begin{tabular}{@{}c@{}}
            Reduce number of gates (\GOne) \quad
            Reduce number of qubits (\GTwo) \quad \\
            Minimize communication (\GThree) \quad
            Gate-set independence (\GFour)
            \end{tabular}
        } \\
        \midrule
        \textbf{Approach} & \textbf{Goals} & \textbf{Approach} & \textbf{Goals} \\
        \midrule
        PyZX~\cite{kissingerPyZXLargeScale2020}                & \GOne & Time-optimal A*~\cite{zhangTimeoptimalQubitMapping2021}             & \GThree \\
        Hybrid ZX-calculus~\cite{borgnaHybridQuantumClassicalCircuit2021}    & \GOne & Layer-by-layer A*~\cite{zulehnerEfficientMappingQuantum2018}             & \GThree \\
        Relaxed peephole optimization~\cite{liuRelaxedPeepholeOptimization2021}         & \GOne & Genetic mapper~\cite{koleImprovedMappingQuantum2020}             & \GThree \\
        Rule-based~\cite{namAutomatedOptimizationLarge2018}          & \GOne & WPM~\cite{siraichiQubitAllocation2018}             & \GThree \\
        QContext~\cite{liuQContextContextAwareDecomposition2023a}   & \GOne  & SABRE~\cite{liTacklingQubitMapping2019}             & \GThree \\
        QSSA~\cite{peduriQSSASSAbasedIR2022}                    & \GOne & Learning-based mapping~\cite{acamporaDeepNeuralNetworks2021}           & \GThree \\
        PCOAST~\cite{paykinPCOASTPauliBasedQuantum2023}          & \GOne & OLSQ~\cite{tanOptimalLayoutSynthesis2020}             & \GThree \\
        Assertion-based optimization~\cite{hanerAssertionbasedOptimizationQuantum2020} & \GOne & SMT-based mapping~\cite{willeMappingQuantumCircuits2019}             & \GThree \\
        SQUARE~\cite{dingSQUAREStrategicQuantum2020}             & \GOne~\GTwo~\GThree & QUEST~\cite{patelQUESTSystematicallyApproximating2022}  & \GOne~\GThree \\
        Unqomp~\cite{paradisUnqompSynthesizingUncomputation2021} & \GOne~\GTwo & VOQC~\cite{hietalaVerifiedOptimizerQuantum2021, hietalaVerifiedOptimizerQuantum2023} & \GOne~\GThree \\
        Reqomp~\cite{paradisReqompSpaceconstrainedUncomputation2022} & \GOne~\GTwo & Queso~\cite{xuSynthesizingQuantumCircuitOptimizers2023} & \GOne~\GFour \\
        QIRO~\cite{ittahQIROStaticSingle2022}                  & \GOne~\GTwo& Parametrized Circuit Inst.~\cite{younisQuantumCircuitOptimization2022} & \GOne~\GFour \\
        Wire Recycling~\cite{palerWireRecyclingQuantum2016}              & \GTwo & Quartz~\cite{xuQuartzSuperoptimizationQuantum2022}       & \GOne~\GFour \\
        Silq~\cite{bichselSilqHighlevelQuantum2020}            & \GTwo & Quarl~\cite{liQuarlLearningBasedQuantum2023}            & \GOne~\GFour \\
        RL-based Mapping~\cite{pozziUsingReinforcementLearning2022}       & \GThree & & \\
        \bottomrule
    \end{tabular}
\end{table}

\subsection{Reducing the Number of Gates (G1)}
\label{sec:g1_reduce_gates}

Reducing the number of gates addresses the noisy hardware challenge (C1) by directly reducing the number of imperfect gates, and indirectly by shortening the time that a qubit is in a superposition state, thus mitigating the effect of decoherence.
A common method to reduce gate count is to apply program transformations, also called rewriting rules, that replace sequences of gates with shorter, equivalent ones.
Most approaches for this goal focus on reducing CNOT gates, as they are frequent and error-prone.
The top circuit in Figure~\ref{fig:sa_optimize_ch_overview_gate_imperfection_and_gate_imperfection} is optimized into the bottom one, reducing gate count and depth.
The example shows a complex optimization done by Quartz~\cite{xuQuartzSuperoptimizationQuantum2022}, which automatically discovers and applies rewriting rules to transform the circuit into an equivalent but shorter version.
Regardless of the techniques, some obstacles in achieving gate reduction are the vast search space for rewrite rules and finding the optimal sequence to apply.
In particular, when some rewrite rules are applied in a sequence some intermediate states might have a higher number of gates than the original state, thus the optimization pass needs to be able to backtrack and try a different sequence of rules.
Moreover, some rewrite rules need a global view of the circuit, i.e., the optimization pass must reason about the entire circuit, not just a local window of gates.
Table~\ref{tab:multiple_optimization_goals} lists approaches targeting gate reduction.

\paragraph{Local program transformations} These approaches apply context-independent transformations, focusing mainly on adjacent gates.
Relaxed peephole optimization~\cite{liuRelaxedPeepholeOptimization2021} uses local optimization methods similar to classical compilers.
It inspects small windows of contiguous operations, replacing them with equivalent sequences having fewer CNOT gates.
User annotations indicating known qubit states, such as basis or pure states, enable more aggressive optimizations.
For instance, if the input state is annotated, it can replace sequences of two-qubit gates, which normally result in three CNOT gates in the Qiskit compiler, with a single CNOT gate and four one-qubit gates using state preparation circuits.
PyZX~\cite{kissingerPyZXLargeScale2020} leverages the theory of the ZX-calculus~\cite{coeckeInteractingQuantumObservables2011} to convert a quantum circuit to a semantically equivalent ZX-diagram representation and then applies rewriting rules to reduce the gates in that intermediate representation.
One limitation is the lack of guaranteed efficient translation of the final ZX-diagram back to a quantum circuit, known as the circuit extraction problem.
Another approach~\cite{namAutomatedOptimizationLarge2018} uses a set of five fixed rules that are applied according to heuristically defined schedules.

\paragraph{Context-aware program transformations} Another class of methods is context-aware, meaning that they also directly or indirectly reason about the context around the circuit segment to optimize. PCOAST~\cite{paykinPCOASTPauliBasedQuantum2023} exploits the commutative properties of the Pauli strings to merge Pauli rotations gates even in the presence of non-unitary operations, such as measurements.
It relies on a Pauli-based representation of the circuit, called \emph{PCOAST graph}, where nodes are both quantum gates and measurements, and edges represent where two gates do not commute, namely their execution order is fixed.
QContext~\cite{liuQContextContextAwareDecomposition2023a} uses a library of possible gate decomposition among which to choose the best variant to decompose Toffoli and CNOT gates depending on the context, namely predecessors and successors of the current gate, and hardware topology.
The gate variant library is created via least square optimization or exhaustive search.
\citet{borgnaHybridQuantumClassicalCircuit2021} extends the ZX-calculus to a hybrid quantum-classical circuit model that includes both quantum and classical operations, and then applies rewriting rules to reduce the number of gates.

\paragraph{Automatically deriving transformations}
Instead of relying on a fixed set of rules, another group of approaches looks for new rewrite rules automatically.
The problem of discovering rules consists in finding a set of circuits, typically small ones, that are equivalent to each other.
Thus, each pair of circuits in the set forms a rewriting rule.
Quartz~\cite{xuQuartzSuperoptimizationQuantum2022} automatically discovers new rewriting rules by using the \emph{RepGen} algorithm that efficiently constructs via recursion all  equivalent classes of circuits with n gates and q qubits, called (n, q)-complete equivalent circuit classes.
To find potentially equivalent circuits, it uses a fast fingerprinting mechanism, followed by a more rigorous and expensive check with an SMT solver.
The new rewriting rules are then applied to the circuits using a cost-based backtracking search algorithm, where the cost is the number of gates in the circuit.
QUESO~\cite{xuSynthesizingQuantumCircuitOptimizers2023} automatically discovers new rewriting rules by using a novel data structure called \emph{polynomial identity filter} to efficiently create clusters of equivalent circuits.
This mechanism converts each circuit into a polynomial and checks if two polynomials derived from different circuits are equal under some randomized evaluation of their variables.
The rewrite rules are then applied with a beam-search algorithm that uses a cost function based on the number of gates in the circuit.
Another work~\cite{meuliSATbasedCNOTQuantum2018}, which we call \emph{SAT-based CNOT}, uses a SAT solver to find the optimal rewrite rules to reduce the number of CNOT gates in all subcircuits that use CNOT and T gates.
They rely on a \emph{phase polynomial representation} that represents each circuit using only CNOT and T gates with a linear reversible function $g$ and a polynomial $p(x_1,...,x_n)$ defining a diagonal phase transformation.
Note that more than one circuit can share the same representation and the goal of the approach is to iteratively find the representation with the smallest number of CNOT gates by solving a SAT problem.

\paragraph{Classical approaches} Some approaches leverage classical optimization techniques to optimize quantum programs.
For example, QIRO~\cite{ittahQIROStaticSingle2022} and QSSA~\cite{peduriQSSASSAbasedIR2022} both leverage the Multi-Level Intermediate Representation (MLIR) to explicitely represent the dataflow in quantum programs via a static single assignment (SSA) graph, but instead of representing flow of classical data, they represent the flow of quantum state from a gate to another.
MLIR allows both to apply classical optimization techniques to quantum programs.
QIRO applies classical optimizations like sub-expression elimination and function inlining, along with quantum-specific ones like gate cancellation and loop optimization. QSSA focuses on quantum-specific optimizations like redundancy and dead-code elimination.

\paragraph{Approximate synthesis}
Approximate synthesis involves replacing a set of gates with a shorter sequence that is close to, but not exactly, the original circuit.
Parametrized Circuit Instantiation~\cite{younisQuantumCircuitOptimization2022} uses a concept called numerical instantiation to replace a usually long sequence of gates with a shorter parametrized circuit that is then optimized in its parameters to find an equivalent circuit via numerical optimization.
Note that the optimization does not guarantee an exact solution.
QUEST~\cite{patelQUESTSystematicallyApproximating2022} splits a circuit into partitions and generates approximate versions of each partition via \emph{approximate synthesis} techniques such that the number of CNOT gates is reduced.
It uses dual annealing optimization to combine partitions and generate a set of diverse versions of the full circuit, such that the output of running the set is both close to the ideal distribution and with lower number of CNOT gates.

\paragraph{Learning-based strategies} Beside knowing the rules to apply, it is crucial to know when to apply them.
For instance, Quarl~\cite{liQuarlLearningBasedQuantum2024} replaces the backtracking search of Quartz with a reinforcement learning agent.
It represents a quantum circuit as a graph and then processes it with a graph neural network.

\subsection{Reducing the Number of Qubits (G2)}
\label{sec:g2_reduce_qubits}

Reducing the number of qubits is a crucial optimization goal to address the constrained resources challenge (C2).
In particular, since most programming languages allow the user to manage memory allocation by allocating quantum memory, the user might sometime use qubits in a suboptimal way, and hence, an optimization can try to reduce the number of qubits used by the program.
The example in Figure~\ref{fig:sa_optimize_ch_overview_constrained_resources} shows a quantum program that uses three qubits that would not fit on a two-qubit quantum computer, whereas the optimized version uses only two qubits.

The example in Figure~\ref{fig:sa_optimize_ch_overview_constrained_resources} is optimized by the approach of \citet{palerWireRecyclingQuantum2016}.
The approach uses the key insight that qubits are usually not needed during the entire computation, but only between their initialization and measurement.
The circuit is represented as a \emph{causal graph}, where nodes are gates, inputs, or outputs.
Edges represent the data-flow dependencies between gates, i.e., the fact that one gate takes the output quantum state of another gate as its input.
\citet{palerWireRecyclingQuantum2016} propose two heuristics to search for the best candidates for wire recycling, based on the pre-defined wire ordering of circuit synthesis methods or the time proximity between two gates in the causal graph.

\paragraph*{Uncomputation}
A significant cluster of approaches~\cite{dingSQUAREStrategicQuantum2020, paradisUnqompSynthesizingUncomputation2021, paradisReqompSpaceconstrainedUncomputation2024, bichselSilqHighlevelQuantum2020} that reduce the number of qubits are based on the idea of \emph{uncomputation}, where the value of the auxiliary qubits that are no longer needed are reverted back to their initial state of $\ket{0}$ by applying the inverse of the operations that were applied to them.
These auxiliary qubits are called \emph{ancilla qubits} and are used to store intermediate results.
Uncomputation is needed in quantum programs because when using ancilla qubits, the quantum state of the ancilla qubits is entangled with the quantum state of the other qubits, and simply discarding them via reset or measurement would also have unwanted side effects on the other qubits.
SQUARE~\cite{dingSQUAREStrategicQuantum2020} combines uncomputation with a cost-benefit analysis to decide which qubits to uncompute.
Unqomp~\cite{paradisUnqompSynthesizingUncomputation2021} avoids adding redundant gates by automatically synthesizing uncomputation blocks whenever there are nested subroutines.
They represent the program as a circuit graph, which is a direct acyclic graph with gates as nodes and dependencies as edges.
Reqomp~\cite{paradisReqompSpaceconstrainedUncomputation2024} extends Unqomp by introducing the concept of recomputation, where instead of uncomputing an ancilla qubit after its last use, it is uncomputed earlier and then recomputed when needed.
This leads to further reduction in the number of qubits used as ancilla.
To achieve this, Reqomp extends the circuit graph representation of Unqomp by adding value indices that track the state of qubits.
Silq~\cite{bichselSilqHighlevelQuantum2020} is a high-level quantum programming language that provides a type system and type checker to better identify uncomputation opportunities.
In particular, Silq introduces the \emph{QFree} and \emph{MFree} function type to annotate functions whose semantics can be described classically and functions without measurements, respectively.
This information is relevant for uncomputation because a \emph{QFree} function introduces no entanglement, making uncomputation easier, whereas knowing that a function is \emph{not} \emph{MFree} means that the function has measurements, which are not reversible and thus the function cannot be automatically reverted by uncomputation.

\subsection{Minimizing Communication (G3)}
\label{sec:g3_minimize_communication}

Current quantum computers typically allow each physical qubit to interact with only a few others (C3).
We refer to the qubit indices used when writing a quantum program as \emph{logical qubits} or \emph{pseudo qubits}, whereas we denote the corresponding qubit indices used on hardware as \emph{physical qubits}.
The problem of finding a suitable correspondence between the logical qubits and the physical qubits is referred to by different names: \emph{qubit mapping}~\cite{zulehnerEfficientMappingQuantum2018}, \emph{placement}~\cite{shafaeiQubitPlacementMinimize2014}, and \emph{allocation}~\cite{siraichiQubitAllocation2018}.
When two qubits need to interact but are not directly connected in hardware, we need to introduce extra gates to move the quantum state from one physical qubit to another until the two logical qubits are physically connected, perform the operation, and then move the quantum state back.
This is done by adding \emph{swap gates} to connect the two states.
This concept of moving quantum information along a specific path of gates is known as \emph{routing}~\cite{cowtanQubitRoutingProblem2019}.
Unfortunately, because each swap gate introduces noise and errors, the optimization pass needs to minimize the number of swap gates inserted, or in other words, minimize the communication cost between the logical qubits.
Figure~\ref{fig:sa_optimize_ch_overview_limited_connectivity} illustrates how a poor mapping can triple the number of swap gates compared to an optimal one.

\begin{figure}[t]
    \centering
    \includegraphics[width=\textwidth]{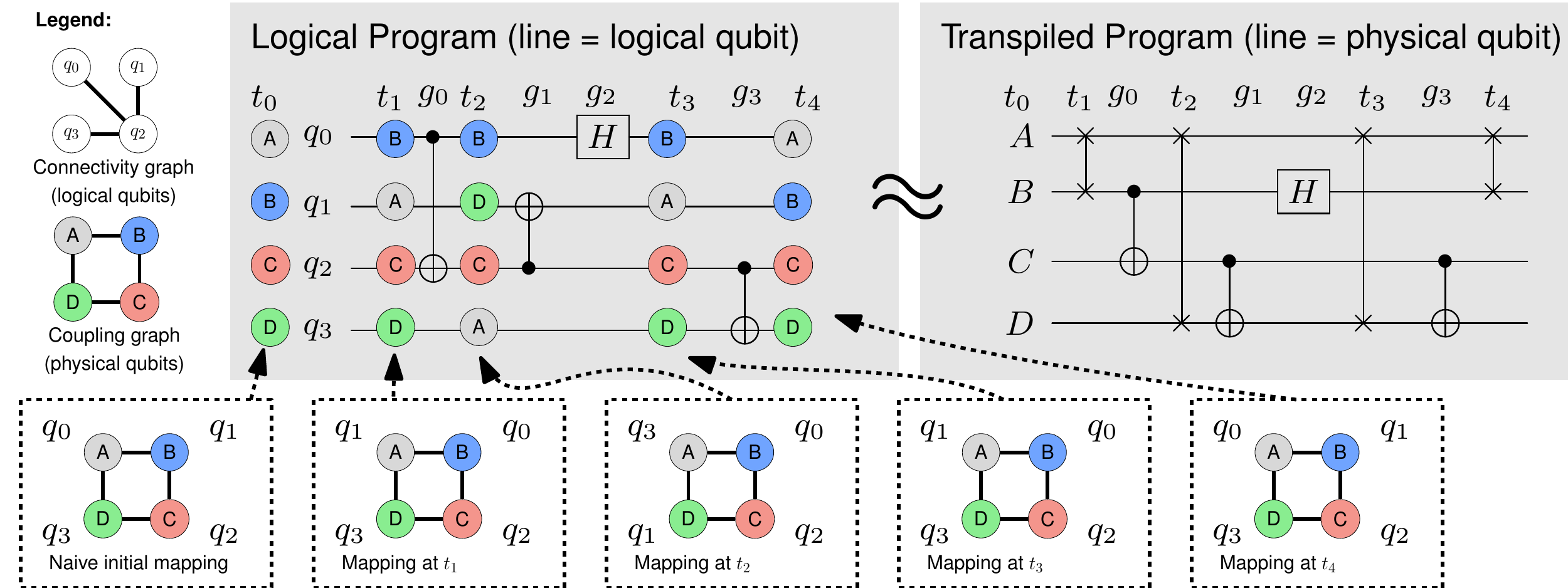}
    \caption{Example of a quantum program's logical view (left) and physical view (right). Hardware constraints necessitate remapping logical qubits to physical ones during computation to ensure all two-qubit operations can be executed on the square coupling graph.}
    \label{fig:sa_optimize_ch_mapping_explanation}
\end{figure}

Figure~\ref{fig:sa_optimize_ch_mapping_explanation} provides a detailed example, showing both the \emph{connectivity graph} (logical qubits and their interactions in the program) and the \emph{coupling graph} (physical qubit connections).
In the top-center, we show the logical program, i.e., the circuit before being mapped to hardware, where each qubit can potentially interact with any other qubits.
However, the initial mapping does not allow the first gate to be performed due to a mismatch between the connectivity graph and the coupling graph.
The first CNOT gate $g_0$ requires qubits $q_0$ and $q_2$ to interact, but in the initial mapping they map to physical qubits $A$ and $C$, which are not connected. Thus, the compiler must adjust the mapping.
This happens at various stages during the computation and each change of mapping corresponds to one or more additional swap gates, as shown in the transpiled program of Figure~\ref{fig:sa_optimize_ch_mapping_explanation}.

The difficulty in achieving this goal is that the qubit mapping problem is NP-complete~\cite{siraichiQubitAllocation2018}, thus the optimization pass needs to rely on heuristics to find a good mapping.
The number of possible mappings or permutations is large and grows factorially ($O(n!)$) with the number $n$ of qubits. In particular, mappings that may initially seem efficient may become inefficient later in the circuit. As a result, the optimization pass needs to efficiently explore the search space.

To mitigate the impact of extra gates introduced by the mapping, various methods aim to minimize added gates or limit circuit depth.
\citet{zulehnerEfficientMappingQuantum2018} and \citet{zhangTimeoptimalQubitMapping2021} both use A* search to find optimal mappings efficiently. They rely on an \emph{admissible heuristic} to guide the search, ensuring the cost (number of swap gates) is always underestimated, thus guaranteeing an optimal solution.
The two approaches~\cite{zulehnerEfficientMappingQuantum2018,zhangTimeoptimalQubitMapping2021} differ in search space exploration.
The former~\cite{zulehnerEfficientMappingQuantum2018} explores layer-by-layer, where a layer is a set of gates executable in parallel with a single mapping.
The latter~\cite{zhangTimeoptimalQubitMapping2021} explores globally, aiming to reduce circuit depth and execution time.

Other approaches use heuristics based on the configuration of the connectivity graph and on the number of pairs of logical qubits interacting via gates.
For instance, \citet{koleImprovedMappingQuantum2020} propose a genetic algorithm to minimize a cost function that includes the distance between two logical qubits in terms of how many swap gates are needed to connect them physically, but also on how many gates are executed between them.
WPM~\cite{siraichiQubitAllocation2018} uses the connectivity graph for initial mapping, offering an exact dynamic programming solution for small circuits and a heuristic for larger ones.

The presence of a gate between two qubits can be seen as a constraint to be considered by a mapping.
Approaches that follow this view apply either satisfiability modulo theories (SMT) solvers~\cite{tanOptimalLayoutSynthesis2020, willeMappingQuantumCircuits2019} or mixed integer programming (MIP)~\cite{shafaeiQubitPlacementMinimize2014}.
The constraints in these mathematical representations encode the connections between physical qubits and between logical qubits.
Once encoded as constraints, the SMT solver or MIP can find the optimal mapping by minimizing the number of swap gates introduced.
However, since solvers are expensive to run, the innovation of each approach is on how they introduce extra constraints to reduce the constraint solving time.
OLSQ~\cite{tanOptimalLayoutSynthesis2020} comes in two variants:
The base OLSQ approach considers all constraints and sends them to the Z3 solver.
The OLSQ-TB approach is more efficient because it reduces redundant mapping variables between so-called transitions, i.e., sets of parallel SWAP gates that are executed between two gates of the original circuit.
\citet{willeMappingQuantumCircuits2019} use a similar approach, but approximates the full SMT problem formulation by noticing that usually only a subset of physical qubits is used by the logical qubits.
Furthermore, their approach introduces swap gates only before certain, heuristically defined gates.
\citet{shafaeiQubitPlacementMinimize2014} formulate the qubit mapping problem as a MIP problem, where the circuit is chunked into layers of $m$ consecutive gates, and then the mapping is optimized for each layer, instead of considering the connectivity graph of the entire circuit.
However, the solution by the MIP solver is not guaranteed to be correct because some logical qubits that should interact might no be close to each other in the final mapping.
Thus, a further check is needed to ensure the correctness of the mapping, followed by occasional re-mapping.

Another way to address the limited connectivity challenge (C3) is via uncomputation, prioritizing uncomputation and reusing nearby qubits.
For example, SQUARE~\cite{dingSQUAREStrategicQuantum2020} introduces a locality-aware allocation strategy to decide which qubit to uncompute based on its costs.
The intuition is that recycling a nearby qubit via uncomputation may be better than moving a fresh qubit from a distant location.
To balance and optimize this trade-off between communication and uncomputation, SQUARE uses a cost-benefit analysis to decide whether to uncompute a qubit or not.

All the previous approaches assume that all physical qubits are identical to each other, and hence, their connections with other qubits are equally reliable.
However, the error rates of applying gates on different physical locations varies.
Some approaches~\cite{tannuNotAllQubits2019, muraliFullstackRealsystemQuantum2019} consider the gate fidelity of different qubits when addressing the qubit mapping problem.
\citet{tannuNotAllQubits2019} propose two schemes called variation-aware
qubit movement (VQM) and variation-aware qubit allocation (VQA) to optimize the movement and allocation of qubits to avoid weak qubits and links.
To do so, their approach collects device characteristics, such as gate fidelities, and uses them to guide the qubit mapping.
Then, when deciding along which path in the coupling graph to move a pair of qubits, VQM will consider the shortest paths and those that are at most $k=4$ nodes longer, to pick the best path with the highest reliability of the links.
Instead, VQA augments the layer-by-layer A*~\cite{zulehnerEfficientMappingQuantum2018} baseline algorithm by computing the most reliable mapping at each layer and by then using the existing A* search to find an optimal way to connect the mapping across layers.
\citet{muraliFullstackRealsystemQuantum2019} show that using an SMT solver that jointly minimizes the number of swap gates and the error rate of the final circuit improves program success rates.
They rely on benchmark data to estimate the error rates of the gates in the hardware.

Another popular heuristic, which is now part of the Qiskit compiler, is SABRE~\cite{liTacklingQubitMapping2019}.
It leverages the insight that, because quantum programs are reversible, finding an optimal mapping for the original program is equivalent to finding an optimal mapping for the reversed program.
SABRE analyzes the circuit in the original order to generate a mapping, and then updates it by analyzing the circuit in the reversed order, thus considering global circuit information.

\paragraph{Learning-based approaches} \citet{acamporaDeepNeuralNetworks2021} propose a learning-based approach to the qubit mapping problem, where they use a deep neural network to predict the best mapping for a given quantum program by learning from a dataset of the best mapping found by the Qiskit transpiler~\cite{acamporaDatasetQuantumCircuit2021}.
The problem is cast as a classification problem, where the input is the quantum program and the output is the best mapping for each qubit.
Since a purely learning-based approach can still generate infeasible solutions, they perform a post-processing step that ensures the generated mapping to be feasible.
\citet{pozziUsingReinforcementLearning2022} use reinforcement learning to find the optimal mapping, where a reward is given each time that two-qubit states are brought close enough to each other in the hardware that a two-qubit gate can be executed.
Analogously to reinforcement learning, \citet{venturelliCompilingQuantumCircuits2018} cast the qubit mapping problem as a temporal planning problem and use a temporal planner to find the optimal mapping.

\subsection{Gate-Set-Independent Optimizations (G4)}
\label{sec:g4_gate_set_independent_optimizations}

As current quantum computers rely on different technologies, the operations they can perform on the qubits may vary.
For example, the gate set used by IBM quantum computers includes the following basic gates:
$U1(\lambda)$, $U2(\phi, \lambda)$, $U3(\theta, \phi, \lambda)$, and the two-qubit CX gate. %
Other vendors and approaches express their optimizations in different gate sets, such as the RzQ gate set used by \citet{hietalaVerifiedOptimizerQuantum2023}, which includes the H, X, $R_{z}Q(q)$, and CX gates. %
To address the challenge of varying gate sets, optimizations ideally should be gate set independent, i.e., able to optimize a quantum program regardless of the gate set being used.
The example in Figure~\ref{fig:sa_optimize_ch_overview_gate_set_variation} shows a pictorial representation of the gate set variation challenge (C4) and its mitigation by being gate set independent and having a hardware-agnostic optimization pass.
The difficulty lies in reasoning about generic unitaries applied to qubits, often involving matrix properties, rather than relying on specific gate properties.
Given the great diversity of gate sets, optimizations that are highly effective for a specific gate set might not be effective for another gate set, thus automatic discovery of new rewrite rules is crucial.

Most of the approaches that are gate set independent have mechanisms to automatically discover new rewrite rules.
Two examples mentioned above are Quartz~\cite{xuQuartzSuperoptimizationQuantum2022} and Queso~\cite{xuSynthesizingQuantumCircuitOptimizers2023}, which automatically discover new rewrite rules by using a fast mechanism to create clusters of equivalent circuits from which rewrite rules can be directly extracted for any gate set.
Another approach that is gate set independent is parametrized circuit instantiation~\cite{younisQuantumCircuitOptimization2022}.
It uses a numerical instantiation technique to find equivalent circuits in the new target gate set.
Unlike previous methods, it does not use rewrite rules but finds optimal parameters to make a circuit in the new gate set equivalent to the original.

\section{Datasets and Benchmarks}
\label{sec:dataset}

This section reviews research efforts providing benchmarks and evaluation settings for testing and analyzing quantum programs. We categorize datasets into general-purpose and task-specific.
Table~\ref{tab:benchmark_dataset} summarizes these datasets and their characteristics.
We cover the following aspects for each dataset: (a) program size, typically in qubits or files; (b) dataset size: number of programs or files; (c) source: collected (e.g., from GitHub) or generated (e.g., from templates); (d) format: language or format used; (e) characteristics: unique features of the programs.
We exclude datasets for quantum machine learning and error correction codes, as they are not directly related to quantum software testing and analysis. We also exclude datasets with only randomly generated circuits without specific applications.

\begin{table}[t]
    \centering
    \caption{Dataset of Quantum Programs. * means dataset not shared by the authors.}
    \label{tab:benchmark_dataset}
    \begin{tabular}{llllll}
      \toprule
      Name & Program size & Dataset size & Source & Format & Characteristics\\
      \midrule
      \multicolumn{6}{l}{\textbf{General-purpose}} \\
      \midrule
      SupermarQ~\cite{tomeshSupermarQScalableQuantum2022a} & 3-1000+ qubits & 52 & Generated & OpenQASM & 8 applications\\
      QasmBench~\cite{liQASMBenchLowLevelQuantum2023} & 433 qubits & 137 & Generated & OpenQASM & 9+ domains\\
      MQT-Bench~\cite{quetschlichMQTBenchBenchmarking2023} & 2-130 qubits & 70,000 & Generated & OpenQASM & 4 abstr. levels\\
      QED-C Benchmark~\cite{lubinskiApplicationOrientedPerformanceBenchmarks2023} & Configurable & 14+ & Generated & Python (Qiskit) & 14 applications\\
      \midrule
      \multicolumn{6}{l}{\textbf{Task-specific: Correctness}} \\
      \midrule
      Bugs4Q~\cite{zhaoBugs4QBenchmarkReal2021} & Single file & 42 & GitHub & Python (Qiskit) & Single platform\\
      \citet{paltenghiBugsQuantumComputing2022} & Multi-file & 223  & GitHub& Multi-language & 18 platforms\\
      \citet{zhaoEmpiricalStudyBugs2023}* & - & 391 & GitHub & Multi-language & 22 frameworks\\
      Muskit~\cite{mendiluzeMuskitMutationAnalysis2022} & Configurable & - & Generated & Python (Qiskit) & -\\
      QMutPy~\cite{fortunatoQMutPyMutationTesting2022} & Configurable & - & Generated & Python (Qiskit) & -\\
      \midrule
      \multicolumn{6}{l}{\textbf{Task-specific: Optimization}} \\
      \midrule
      RevLib~\cite{willeRevLibOnlineResource2008} & Configurable & 154+ & Generated & Custom format & 8 families\\
      Queko~\cite{tanOptimalityStudyExisting2021a} & Up to 900 qubits & 900+ & Generated & OpenQASM & Mapping stage\\
      \citet{acamporaDatasetQuantumCircuit2021} & 2-15 qubits & 47,111 & Generated & Connectivity Graph & Calibration data\\
      \citet{moriQuantumCircuitUnoptimization2023} & Configurable & - & Generated & Python (Qiskit) & -\\
      \bottomrule
    \end{tabular}
\end{table}

\subsection{General-Purpose Datasets}
By general-purpose dataset, we mean a collection of diverse quantum programs without task-specific details or ground truth.
SupermarQ~\cite{tomeshSupermarQScalableQuantum2022a} includes 52 circuits ranging from 3 to 1000+ qubits.
Generated using templates from eight benchmark applications, these programs cover diverse features measured by metrics like entanglement ratio, critical depth, and liveness.
Each program has a known solution or is efficiently simulatable on classical computers.
QASMBench~\cite{liQASMBenchLowLevelQuantum2023} includes 48 circuits of varying sizes, divided into small (2-5 qubits), medium (6-15 qubits), and large (15+ qubits). The latest version has 137 circuits, with the largest at 433 qubits.
The dataset has been used to evaluate testing tools like Giallar~\cite{taoGiallarPushbuttonVerification2022} and analysis tools like QSSA~\cite{peduriQSSASSAbasedIR2022}.
MQT Bench~\cite{quetschlichMQTBenchBenchmarking2023} includes 70,000 circuits ranging from 2 to 130 qubits.
It supports benchmarking at four abstraction levels: algorithmic, target-independent, target-dependent with native gates, and target-dependent with specific connectivity graphs.
Programs are in OpenQASM, generated from templates.

The QED-C Benchmark~\cite{lubinskiApplicationOrientedPerformanceBenchmarks2023} contains application-oriented benchmark programs divided in three categories: tutorial, subroutines, and functional.
Each benchmark program is scalable, i.e., representing a family of circuits that can be automatically generated.
At the time of writing, the official repository contains 14 benchmark circuits where the number of qubits is configurable up to what is supported by the simulator.
The programs are written in Qiskit, Cirq, and Braket.
However, only the Qiskit version is available for all the benchmark circuits.
Although mainly designed to benchmark hardware, the benchmark could also be used with testing and analysis techniques.

\subsection{Task-Specific Datasets}

These datasets either contain known bugs to evaluate the correctness of testing tools,  and bug detectors, or they are used to validate the effectiveness of analysis tools, such as mapping or optimization tools.

\paragraph*{Correctness Benchmarks} Benchmarks targeted at the correctness of quantum programs typically are programs with known bugs, and optionally their corresponding fixes.
Often, these programs are extracted from the version history of quantum computing repositories.
Regarding the correctness of quantum circuits, Bugs4Q~\cite{zhaoBugs4QBenchmarkReal2021} contains 42 buggy programs written in Qiskit.
They are sourced from developers' discussions on GitHub issues and StackOverflow, and each is accompanied by a corresponding bug fix.
The programs are small circuits or excerpts of larger programs.
For example, LintQ~\cite{paltenghiAnalyzingQuantumPrograms2024} uses this dataset to assess the recall of its static analysis in identifying programming issues.
Regarding the software of the platforms, \citet{paltenghiBugsQuantumComputing2022} collected 223 buggy program from the version history of 18 quantum computing platforms.
Each bug is accompanied by a corresponding minimal bug fix including multi-file fixes.
Analogously \citet{zhaoEmpiricalStudyBugs2023} collected 391 real-world bugs, but from 22 quantum machine learning frameworks, such as PennyLane and TorchQuantum. The dataset is not publicly available, however.
Besides collecting bugs, other approaches generate mutants of quantum circuits to evaluate the effectiveness of testing tools in detecting bugs.
Muskit~\cite{mendiluzeMuskitMutationAnalysis2022} generates mutants of quantum circuits using mutation operators like gate replacement, insertion, and deletion.
QMutPy~\cite{fortunatoQMutPyMutationTesting2022} extends the same idea by ensuring replaced gates match the original gate's input qubits.
Both tools operate on Qiskit programs and can generate numerous mutants.
\citet{usandizagaWhichQuantumCircuit2023} use Muskit to study mutant characteristics, releasing a dataset of 700,000 mutants from 382 real-world circuits.

\paragraph*{Optimization Benchmarks} Another popular task-oriented family of datasets is used to evaluate techniques that optimize quantum programs (Section~\ref{sec:static_analysis_optimize}).
RevLib~\cite{willeRevLibOnlineResource2008} contains reversible circuits to benchmark reversible logic synthesis tools.
It contains 154 functions for generating reversible circuits, each serving as a template for a family of circuits with different sizes.
The circuits are grouped into eight families, one of which is specifically for quantum gates and contains five functions.
The programs include adders, multipliers, and other arithmetic circuits.
The format is a custom format that describes the circuit as a list of gates.
QUEKO~\cite{tanOptimalityStudyExisting2021a} is a dataset of quantum circuits in the OpenQASM format, to be used in the mapping stage of quantum circuit compilation.
The authors create the dataset by heuristically generating circuits where the optimal layout is known.
The underlying circuits are designed to run on coupling graphs with up to 54 qubits.
The benchmark is divided into two parts: one with circuits of low depth (up to 45) for NISQ devices, and one with high depth (up to 900) for scaling studies.
\citet{acamporaDatasetQuantumCircuit2021} introduce a dataset of random quantum circuits with their corresponding best layout found by the Qiskit transpiler when given a specific calibration data as input.
The circuits have sizes ranging from 2 to 15 qubits, and the dataset contains 47,111 circuits.
Note that only the connectivity map of each circuit is provided, not the circuit itself.
This dataset is used to evaluate a learning-based approach to find the best layout for a given quantum circuit~\cite{acamporaDeepNeuralNetworks2021}.
Similar to artificially generating bugs to evaluate testing tools for correctness, \citet{moriQuantumCircuitUnoptimization2023} propose an unoptimization approach to generate quantum circuits as a benchmark for optimization tools.
The approach uses four operations: redundant gate insertion, commuting gate swapping, gate decomposition, and gate synthesis. Programs are generated in Qiskit and PyTket, and evaluated on respective platforms.

\section{Outlook and Challenges for Future Research}
\label{sec:outlook}

Although initial progress has been made in the area of quantum software testing and analysis, the problems it deals with are far from being solved.
This section discusses open research challenges for future work to address.

\subsection{Scalability: Efficient Analysis and Optimization Techniques}
Most existing analyses techniques are applied to relatively small programs, e.g., with few dozens of qubits.
However, as the size of available quantum computers is increasing, we expect quantum software to follow a similar trend.
In response, testing and analysis techniques will have to scale to larger programs.
Some ideas to address this challenge include the use of machine learning techniques to improve the scalability of optimization methods, such as efficiently using reinforcement learning to schedule and optimize quantum operations~\cite{pozziUsingReinforcementLearning2022, venturelliCompilingQuantumCircuits2018}.
Alternatively, high performance computing methods can be incorporated in new analyses methods to capture the behavior of large programs in an efficient but precise way.
In particular, results from high-performance computing for efficient quantum simulation approaches could inspire the software analysis community on how to scale their approaches.
Another promising direction is to customize analyses and optimizations to specific program classes, such as optimizing for a particular circuit family to enhance efficiency.

Beyond the scalability of testing and analysis techniques, future research also will have to increase the scale of evaluations and benchmarks toward programs with hundreds or thousands of qubits.
Some existing benchmarks, e.g., at the intersection with hardware benchmarking~\cite{quetschlichMQTBenchBenchmarking2023,lubinskiApplicationOrientedPerformanceBenchmarks2023}, generate large-scale quantum programs from templates, which provides a first step toward larger-scale benchmarks.
A possible bottleneck in scaling up benchmark size is the need of ground truth data to evaluate the effectiveness of the testing and analysis techniques.
In particular to benchmark mapping approaches (Section~\ref{sec:g3_minimize_communication}) an optimal mapping needs to be known to evaluate the quality of the mapping found by a novel approach.
Similarly, for bug detection (Section~\ref{sec:pattern_based}) a set of known bugs is needed to evaluate the effectiveness of the tool.
Some possible directions include the automatic generation of ground truth data, e.g., by using unoptimization~\cite{moriQuantumCircuitUnoptimization2023} or large language models that inject realistic bugs into existing quantum programs.

\subsection{Quantum-Specific Test Oracles}
Test oracles for quantum software can either adapt ideas from classical computing, such as looking for program crashes, or leverage the probabilistic nature of quantum computing.
However, early attempts of using statistical tests to achieve the latter~\cite{wangQDiffDifferentialTesting2021} turn out to be ineffective when paired with an automatic test generation approach, such as fuzzing.
The reason is that the probability that a statistical oracle reports a false positive increases when larger numbers of tests are generated and executed.
Future work could further study the effectiveness of different statistical tests to understand which tests are suitable as a high-precision test oracle.
Another direction  to alleviate the oracle problem could be to leverage methods from the equivalence checking domain~\cite{pehamEquivalenceCheckingQuantum2022}. However, the scalability of these methods remains an open question.
Simulators have proven beneficial for testing quantum platforms~\cite{paltenghiMorphQMetamorphicTesting2023,xiaFuzz4AllUniversalFuzzing2024}.
However, extending their use to large programs is challenging due to the exponential space required for perfect simulation.
Abstraction-based analysis approaches (Section~\ref{sec:abstraction_based_automation}) offer a first step to address this challenge by leveraging approximate state reasoning.
More work is needed to obtain a reliable testing oracle out of those methods, which often still rely on developer specifications.

\subsection{Balancing Standardization and Diversity}
OpenQASM~\cite{crossOpenQuantumAssembly2017,liQASMBenchLowLevelQuantum2023} and Qiskit are becoming de facto standards in the quantum computing field, a trend that has proven beneficial for interoperability and tool adoption.
Their widespread adoption by the software testing and analysis community is helping to solidify the real-world impact of new testing and analysis methodologies.
However, as more and more techniques and datasets primarily utilize QASM or Qiskit, there is a growing concern about the loss of diversity and the ability to generalize findings.
We envision future work to strike a balance between embracing these standards and exploring new languages and platforms.
Some emerging trends are the potential reuse of classical compiler infrastructure and intermediate representations bridging to the LLVM and MLIR ecosystems~\cite{ittahQIROStaticSingle2022, peduriQSSASSAbasedIR2022}.

\subsection{Developer Tools for Debugging}

Understanding and debugging a quantum program is still significantly more difficult than, e.g., a classical Java or Python program.
Important reasons are the low level at which quantum programs are written and the unintuitive quantum mechanical properties.
Future work on debugging techniques could aim for intuitive program state visualizations and interactive debugging tools to support the developers.
As seen in Section~\ref{sec:testing}, providing a step-by-step execution of a quantum program largely remains an open challenge.
Despite some initial progress~\cite{metwalliToolDebuggingQuantum2022}, more work is needed, including work that seeks validation from developers regarding the usefulness of a debugging tool.
A promising direction is exploiting results from the quantum error correction community for debugging, as illustrated by the work of \citet{liuQuantumCircuitsDynamic2020}.
The testing and debugging community could also include quantum developers in the design and evaluation of these tools, as their feedback will certainly help in understanding their most pressing pain-points.

\subsection{Large Language Models for Quantum Software Testing}

Large Language Models (LLMs) have shown significant potential in generating and reasoning about source code in classical computing.
Fuzz4All~\cite{xiaFuzz4AllUniversalFuzzing2024} represents a successful application of LLMs in quantum platform testing, where an LLM is used to generate valid quantum programs for testing the Qiskit platform.
The abilities of LLMs in understanding recurrent patterns in quantum computing code could be used for other testing and analysis tasks.
For instance, LLMs could automatically infer annotations and specifications for quantum programs, e.g., by annotating which qubit serves as ancilla, which can lower the barrier for adopting formal methods (Section~\ref{sec:formal_methods_abstraction}).
Another promising direction is using LLMs to find effective heuristics and code variants for optimizing quantum programs.
Finally, LLMs could automatically fix bugs detected by static analysis tools (Section~\ref{sec:pattern_based}), thus automating program repair~\cite{gouesAutomatedProgramRepair2019}.

\subsection{Raising the Abstraction Level}

As quantum computing evolves toward more complex programs, we expect to see programming models with high-level abstractions.
Currently, quantum computing programs are predominantly defined at the gate level, which can be cumbersome and limit higher-level conceptual thinking.
Initiatives like OpenQASM~3~\cite{liQASMBenchLowLevelQuantum2023} and ScaffCC's High Level QASM~\cite{javadiabhariScaffCCFrameworkCompilation2014} have begun addressing these challenges by introducing more abstract programming primitives.
Such abstractions demand for future developments in quantum software testing and analysis to understand the higher-level structures and patterns emerging in quantum algorithms.
As a concrete example, future mutation testing techniques could evolve beyond simple gate-level mutations~\cite{fortunatoQMutPyMutationTesting2022} by offering more complex mutations, such as altering quantum oracle specifications, removing specific quantum calls, or recomposing circuits in novel ways.
Similarly, static analysis methods (Section~\ref{sec:pattern_based}), such as LintQ~\cite{paltenghiAnalyzingQuantumPrograms2024}, which currently operate largely at the gate level, must advance to understand and reason about more complex code constructs.

\subsection{Correctness Specification for Dynamic Circuits}
Most of the current quantum programs have a fixed number of operations known at compile time, followed by measurement.
Recent advance in quantum hardware allow for a new class of programs where this is not always true.
These programs, called \emph{dynamic circuits}, allow for operations that depend on the measurement outcome of previous operations or on the result of some concurrent classical computation.
Note that the classical computation can be done on a classical computer, but its result needs to be fed into the quantum computer at runtime so that the quantum program can use it and decide which operation to execute next.
An example of dynamic circuit are the quantum neural networks~\cite{schuldMachineLearningQuantum2021} and the quantum variational algorithms~\cite{cerezoVariationalQuantumAlgorithms2021}, where the quantum program is a sequence of operations, which are interleaved by measurement and classical optimization routines, and those operations are iteratively executed until a certain condition is met.
Program using the new hardware features open up new challenges for program analysis, e.g., when reasoning about the control flow of a program.
Moreover, this new class of programs will likely need novel kinds of specification of the expected behavior, since a single output distribution is insufficient in describing the behavior across iterations.
Some idea could involve the use of multiple admissible output distributions, one for each iteration of the algorithm, or a theoretical framework that reasons on the changes in output distributions between iterations.

\section{Conclusion}
In this survey, we give a unified overview of the critical challenges surrounding testing and analysis of quantum software.
We discuss the state of the art in the field and provide an extensive overview of the existing literature, spanning several research communities, including quantum computing, software engineering, programming languages, and formal methods.
We discuss how the testing problem is formulated by considering both expected and unexpected behaviors of quantum programs.
We report on static analysis methods for both correctness and optimization of quantum programs.
We also provide an overview of existing datasets and benchmarks that can support future research.
Finally, we highlight open challenges in the field with the hope of inspiring new work and advancements.

\section{Acknowledgments}

This work was supported by the European Research Council (ERC, grant
agreements 851895 and 101155832), and by the German Research Foundation
within the ConcSys, DeMoCo, and QPTest projects.

\bibliographystyle{ACM-Reference-Format}
\bibliography{Survey-QuantumTestingAndAnalysis,moreRefs}

\end{document}